\documentclass{article}

\newtheorem{theorem}{Theorem}
\newtheorem{defn}{Definition}
\newtheorem{lemma}{Lemma}
\newtheorem{cor}{Corollary}

\newcommand\norm[1]{\left\lVert#1\right\rVert}

\usepackage{blindtext}
\usepackage{arxiv}
\usepackage{amsmath}
\usepackage{mathtools}
\usepackage[utf8]{inputenc} 
\usepackage[T1]{fontenc}    
\usepackage{hyperref}       
\usepackage{url}            
\usepackage{booktabs}       
\usepackage{amsfonts}       
\usepackage{nicefrac}       
\usepackage{microtype}      
\usepackage{lipsum}
\usepackage{graphicx}
\usepackage{subcaption}
\usepackage{bbm}
\usepackage{cancel}
\usepackage{amssymb}

\graphicspath{ {./images/} }

\title{Modelling heterogeneous outcomes in multi-agent systems}

\author{
Orowa Sikder \\
  Department of Computer Science\\
  University College London\\
  London, WC1E 6BT \\
  \texttt{orowa.sikder.15@ucl.ac.uk}
}

\begin{document}
\maketitle
\begin{abstract}
A broad set of empirical phenomenon in the study of social, economic and machine behaviour can be modelled as complex systems with averaging dynamics. However many of these models naturally result in consensus or consensus-like outcomes. In reality, empirical phenomenon rarely converge to these and instead are characterized by rich, persistent variation in the agent states. Such heterogeneous outcomes are a natural consequence of a number of models that incorporate external perturbation to the otherwise convex dynamics of the agents. The purpose of this paper is to formalize the notion of heterogeneity and demonstrate which classes of models are able to achieve it as an outcome, and therefore are better suited to modelling important empirical questions. We do so by determining how the topology of (time-varying) interaction networks restrict the space of possible steady-state outcomes for agents, and how this is related to the study of random walks on graphs. We consider a number of intentionally diverse examples to demonstrate how the results can be applied.
\end{abstract}


\section{Introduction}


Many empirical phenomenon in the study of complex systems can be modelled as a form of averaging dynamics. In such models, we are interested in modelling the dynamic and steady-state behaviour of a large set of interacting agents, where a key feature is that agents tend to move closer to the states of their neighbours over time. For example, in the study of social behaviour, such models are often employed in the field of social learning or opinion dynamics, where agents each possess a state expressing their opinion, and agent's opinions will gravitate towards the opinions of other agents they interact with regularly (such as DeGroot \cite{degroot1974reaching} or bounded confidence models \cite{hegselmann2002opinion,weisbuch2003interacting}). In studies of economic behaviour, such models arise naturally in the study of strategic complements or co-ordination games, and are closely related to the theory of linear-quadratic games \cite{jackson2015games}. Averaging dynamics also play a key role in the study of animal behaviour in the form of swarm dynamics \cite{vicsek1995novel}, and more recently have been applied to understand some elements of machine behaviour \cite{rahwan2019machine} such as the feedback loop between recommender systems and user preferences \cite{sikder2020minimalistic}.

Such models of averaging dynamics overlap in an important way with control theory and the design of algorithms that can distribute computation over decentralised agents. In an important pair of reviews \cite{proskurnikov2017tutorial,proskurnikov2018tutorial}, the authors synthesize a rich set of theoretical results from across opinion dynamics and control theory to demonstrate this correspondence. In particular, the reviews discuss in detail the conditions under which we expect classes of models to converge, and if they do converge, whether it results in a consensus where all agents achieve the same state, or near-consensus outcomes where subsets of agents all achieve the same state.

While the convergence of such models to consensus is certainly a desirable outcome in the design of algorithms and control systems, in general this is not what we would like when we are modelling real empirical phenomenon, which are often characterised by what we might refer to informally as ``rich'' outcomes: persistent individual variation which tends towards a continuum of outcomes as the size of the models grow, allowing for complex distributions to be realised in the steady state. We refer to these as \textit{heterogeneous outcomes}, with a formal definition to follow (see Figure \ref{fig:hkmodel} for a visual depiction). Indeed, in a review of social learning models \cite{golub2017learning}, the author concludes ``Long-run consensus is a central finding throughout this literature, occurring for a wide range of information structures and decision rules in large classes of networks...The consistency of this finding may cause some discomfort because we often observe disagreement empirically, even about matters of fact. Explaining such disagreement is an important task for this literature going forward.''

An important class of averaging dynamics models that \textit{do} tend to provide us with these heterogeneous outcomes are those models that include ``stubborn agents'' \cite{ghaderi2014opinion} or ``zealots'' \cite{masuda2015opinion}, which influence the dynamics of other agents but are not influenced themselves. Another class which provides these outcomes are Friedkin-Johnsen models \cite{friedkin1990social} and its time-varying alternative \cite{proskurnikov2017opinion}, which are characterised by ``prejudiced'' agents whose dynamics are perturbed by an external vector referred to as an agent's prejudice (in the original models an agent's prejudice was their state at time $0$, demonstrating a form of hysteresis).

Clearly, there exists some correspondence between these classes of models that allow for the modelling of heterogeneous outcomes, and one might speculate that it is intimately related to the connectivity of the (potentially time-varying) interaction network between agents. The objective of this paper is to formalise this notion, and provide a set of necessary criteria for our model of averaging dynamics to result in the heterogeneous outcomes. As such, we hope to provide utility for future research endeavours and support the explicit construction of models that are better suited to real world phenomenon.

In order to do this we proceed in the following manner. We first introduce a ``generalised'' model of averaging dynamics that can realise as a special case a number of different existing influential models. We then demonstrate this general class is isomorphic to an ``augmented'' graph representation that allows us to explicitly tie the steady state outcomes to graph-theoretic features. We provide a number of Theorems that develop necessary conditions for heterogeneity across our broad class of models. We also provide a conceptual bridge between these models and the study of random walks on graphs, and show how such a framework can provide a valuable shortcut to evaluate how topological features of the graph can dictate the distributional outcomes for the agents. Finally, we consider an intentionally diverse set of examples, where we demonstrate the application of different sets of results to models in social, animal, machine and economic behaviour.

We believe the key contributions of this paper are threefold. Firstly, we provide a general and intuitive framework of averaging dynamics in multi-agent systems that contains as a special case many important and influential models in the field. Secondly, we introduce a formal notion of heterogeneity and establish the necessary conditions to achieve these outcomes in the steady state, through a modest generalisation of existing results in the convergence of such models. Finally, we establish conditions for the convergence of these heterogeneous models and provide an intuitive framework to understand such models in the theory of random walks on graphs.

\section{Model and Definitions}

Consider a set of agents $\mathcal{V} = \{v_1, v_2, \ldots, v_N\}$ where each agent $i$ possesses a state $x_i \in X$ where $X$ is some compact subset of $\mathbb{R}^d$. The states of all agents can be represented by $x = (x_1, x_2, \ldots, x_N) \in X^N = \mathcal{X} \subset \mathbb{R}^{N \times d}$. For this paper, we will be interested in the class of quasi-linear update dynamics that can be represented as:

\begin{equation}\label{eq:affine}
    x(t+1) = (I-\Lambda) A(t) x(t) + \Lambda b(t)
\end{equation}

Where the adjacency matrix $A(t)$ is a row stochastic matrix implicitly representing interactions between agents, $b(t) \in \mathcal{X}$ is a private signal associated with each agent, and $\Lambda$ is a diagonal matrix with $\Lambda_{ii} = \lambda_i \in [0,1)$ a parameter that weights the influence of each update component for the $i$-th node. The greater $\lambda_i$, the more weight each agent places on the set of private signals, and the less weight they place on ``social'' signals\footnote{The results in this paper can also be easily generalised to the case where $\Lambda = \Lambda(t)$, but necessitates a great deal of extra notation. We assume the fixed $\Lambda$ for simplicity.}. Note that the framework is not very restrictive, as $A(t) = A(x(t)) = f(x(t))$ can represent any function $f$ of $x(t)$ such that $f_i(x(t))$ can be expressed as a convex combination over the states $x$ of agents at time $t$. Similarly, $b(t) = b(x(t))$ can be quite broad, representing any function $g: \mathcal{X} \to \mathcal{X}$.

This general class of update dynamics contains as a special case a number of different existing and influential models. For example, traditional DeGroot updating \cite{degroot1974reaching} can be retrieved for $\Lambda = 0$, $A(t) = A$. Similarly, bounded confidence models \cite{lorenz2007continuous} can also be characterised by $\Lambda = 0$, but $A(t) = A(x(t))$, where the weights between nodes is determined by whether two agents are within the threshold required to interact. By allowing for $\Lambda > 0$, we include models that incorporate private signals for each agent. For example, for $\Lambda > 0$, $A(t) = A$ and $b(t) = b$, we return to the standard Friedkin-Johnsen model \cite{friedkin1990social}, and allowing $A(t)$ to vary gives us its time-varying alternative \cite{proskurnikov2017opinion}. If $b(t)$ is drawn at each time step from some distribution $\rho_b$ over $\mathcal{X}$ we can model noisy averaging processes. Finally, if $b(t) = b(x(t))$ we can model processes where the private signals received by agents are endogenous and incorporate feedback loops, as the authors previously explored in \cite{sikder2020minimalistic}.


In order to develop properties for this general class of models, it is useful to consider an alternative ``augmented'' representation of original affine form:

\begin{equation}\label{eq:augmented}
    \tilde{x}(t+1) = \tilde{A}(t)\tilde{x}(t)
\end{equation}

Where:

\begin{align}\label{eq:block}
  \tilde{A}(t) = \left[
        \begin{array}{c | c}
        (I-\Lambda) A(t) & \Lambda W(t)\\
        \hline
        0 & I
        \end{array}
        \right] \\
    \tilde{x}(t) =  \left[
        \begin{array}{c}
        x(t) \\
        \hline
        C
        \end{array}
        \right]
\end{align}

        


\begin{figure}
    \centering
    \includegraphics[width=\textwidth]{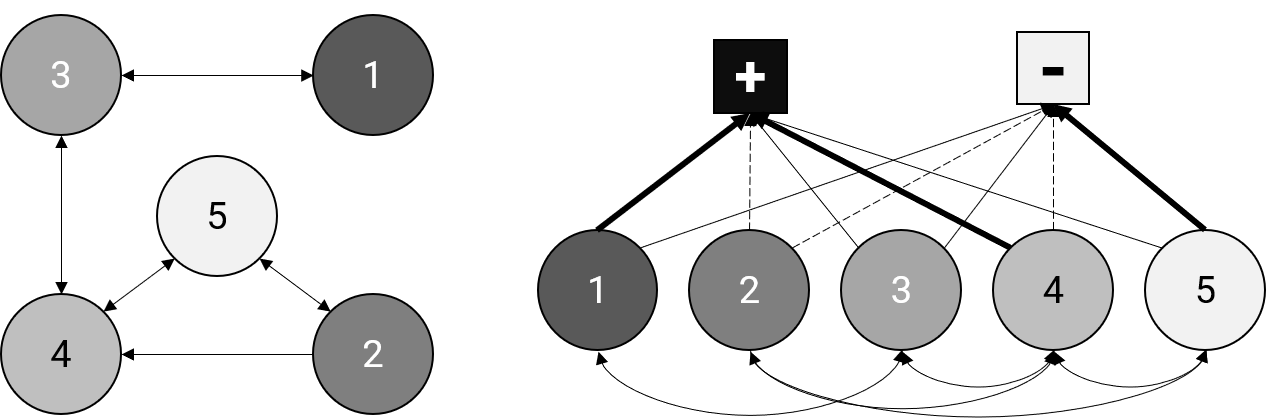}
    \caption{Illustrating how an original graph $\mathcal{G}$ with private signals (left) can be re-formulated as an augmented graph $\tilde{\mathcal{G}}$ (right). The state space of the original nodes is $[-1,1]^N$, so only two ghost nodes are required, one with value $+1$ and one with value $-1$. Nodes with darker shade have a more positive steady state, and we can see on the augmented graph that nodes with a heavier weight on the positive ghost node tend to have a more positive steady state. An alternative formulation would provide \textit{each} original node with its own pair of ghost nodes.}
    \label{fig:ghosts}
\end{figure}
        
Here $\tilde{A}(t)$ is a row stochastic matrix that includes as blocks the original update matrix $A(t)$ as well as a \textit{weight} matrix $W(t)$, where $W(t)C = b(t)$. The block $C$ includes the upper and lower bounds of each dimension in $X$, meaning that any $b(t) \in \mathcal{X}$ can be represented with an appropriate choice of weights $W(t)$. One way of interpreting this augmented representation is to consider a set of ``ghost nodes'' associated with each of the original agents that represent the upper and lower bounds of $X$. The ``private signals'' can be re-interpreted on the weight each agent places on their ghost nodes. More technical details of this augmented structure are provided in the Appendix.

The convenience of the augmented framework in Equation (\ref{eq:augmented}) is that we can ignore the ``augmented'' nature of the dynamics and just consider the general properties of (time-varying) linear updates on a modified graph, without having to make specific considerations for the effect of private signals. The topology of the graph in question will dictate specific features of the dynamics. For example, if there are no private signals in our model (i.e. $\Lambda = 0$), then the augmented matrix $\tilde{A}(t)$ will simply have a block diagonal structure, and the graph it implicitly represents will have ghost nodes disconnected from the original set of nodes. On the other hand if the private signals exist but vary over time then the edge weights to the corresponding ghost nodes change over time. Regardless of the underlying dynamics we choose, the states of the agents at any time step $t$ can be represented compactly as:

\begin{equation}
    \tilde{x}(t) = \prod_{k=0}^t \tilde{A}(k) \tilde{x}(0) = \tilde{A}(t:0) \tilde{x}(0)
\end{equation}

Where the notation $M(t_1:t_0)$ represents the product of matrices $M(k)$ from $t_0$ to $t_1$.

Therefore until stated otherwise, we ignore the special structure of the affine updates with private signals, and simply consider a generic set of nodes $\mathcal{V}$ with states $x(t)$ and a general linear update $A(t)$:

\begin{equation}\label{eq:main}
    x(t+1) = A(t) x(t)
\end{equation}

By analyzing properties of this generic update, we can draw important conclusions about the general class of dynamics expressed in Equation \ref{eq:affine}.


\subsection{Heterogeneous outcomes}

We are interested in analyzing the steady state outcome $x^*$ of such averaging dynamics, assuming they exist. The following definitions are useful in this regard:

\begin{defn}

    (Convergence). A model \textit{converges} iff for any initial state $x(0)$, the limit $x^* = \underset{t\to\infty}{\lim}x(t)$ exists.

\end{defn}

\begin{defn}

    (Consensus). A convergent model reaches \textit{consensus} iff $\underset{t\to\infty}{\lim}x_i(t) = \underset{t\to\infty}{\lim}\bar{x}(t)$ for all $i \in \mathcal{V}$.

\end{defn}

We can see that a set of dynamics that result in consensus will have each agent possess the same steady state asymptotically. It turns out that this is characteristic of a broad class of models \cite{golub2017learning}, even though the vast majority of real phenomenon do not exhibit this. Persistent individual variation across agents is a key empirical feature we want to be able to capture, and in fact has been flagged as a shortcoming of models that always result in consensus \cite{golub2017learning}. In order to characterise non-consensus outcomes, consider the following:

\begin{defn}\label{defn:hetero}

    (Heterogeneity). Define the heterogeneity of $x$ as:
    
    \begin{equation}
        \mathcal{H}(x) = \underset{i,j}{\min} |x_i - x_j|
    \end{equation}
    
    A state $x$ is \textit{heterogeneous} iff $\mathcal{H}(x) > 0$.
    
\end{defn}

\begin{defn}\label{defn:frag}

    (Fragmentation). A non-consensus state $x$ is fragmented iff $\mathcal{H}(x) = 0$.
    
\end{defn}

\begin{figure}
        \centering
        \begin{subfigure}[b]{0.475\textwidth}
            \centering
            \includegraphics[width=0.6\textwidth]{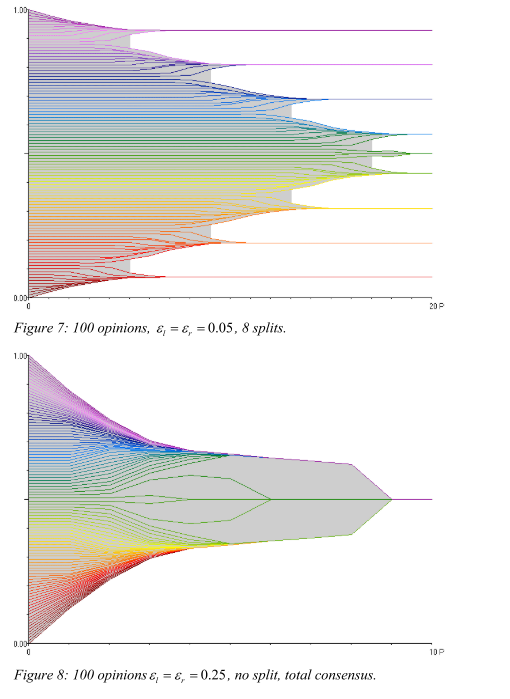}
            \caption[Contrarian agents at $t=10$]%
            {{\small Non-heterogeneous outcomes in bounded confidence models.}}
        \end{subfigure}
        \hfill
        \begin{subfigure}[b]{0.475\textwidth}  
            \centering
            \includegraphics[width=\textwidth]{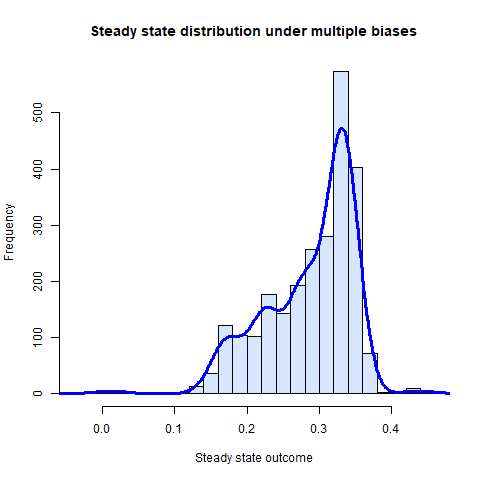}
            \caption[Contrarian agents at $t=20$]%
            {{\small Heterogeneous outcomes in the biased learner model.}}
        \end{subfigure}
    \caption{(Left) A typical example of non-heterogeneous outcomes: Hegselmann-Krause dynamics for the one-dimensional case for two different values of $\epsilon$, resulting in fragmented (top) and consensus (bottom) outcomes in the asymptotic distribution. Taken from \cite{hegselmann2002opinion}. (Right) A heterogeneous outcome as seen in the biased learner model, where agents converge to a continuum of steady state outcomes instead. Histogram indicates steady state outcomes of numerical simulations and solid lines indicate a kernel density estimate, which can be shown to approximate the continuum outcome as $N\to\infty$ \cite{sikder2020minimalistic}.}
    \label{fig:hkmodel}
\end{figure}

Intuitively, fragmented states are those where multiple steady state outcomes exist, but multiple agents still converge to the same outcome. A typical example of this are the outcomes that result from bounded confidence models, where for small enough thresholds $\epsilon$, the agents will fragment into a set of $m$ outcomes, where typically $m \ll n$. While this is certainly a step in the right direction with regards to modelling ``rich'' outcomes for the agents, we still might suppose that outcomes should approach a continuum of steady states as $n$ grows large. The notion of heterogeneity captures this intuition. Heterogeneous outcomes are precisely those with persistent individual variation: each agent possesses a unique steady state outcome. An example is provided in Figure \ref{fig:hkmodel}, where we contrast the fragmented outcome of the dynamics of a bounded confidence model with the distributional outcomes of a biased learner model.

\section{Consensus, convergence and heterogeneity}

We now show that the topology of the graph implicitly represented by the linear updates $A(t)$ can dictate some important properties of the steady states that can be generated. In order to do so we need to be able to characterise time-varying interactions, and do so using a long-run interaction graph, as in \cite{proskurnikov2018tutorial}. We define the infinite graph $\mathcal{G}_{\infty} = \{\mathcal{V}, \mathcal{E}_{\infty}\}$, where $(i,j) \in \mathcal{E}_{\infty} \iff \sum_t A_{ij}(t) = \infty$. That is, if an edge exists between $i$ and $j$, it implies they interact indefinitely over the course of the dynamics.



An important result in \cite{lorenz2005stabilization,blondel2005convergence} and summarised in \cite{proskurnikov2018tutorial} offers sufficient conditions to characterise steady state outcomes. In particular, suppose there exists some $t_0$ such that for all $t \geq t_0$ the following assumptions hold for $t \geq t_0$:

\begin{enumerate}
    \item $A_{ij}(t) \in \{0\} \cup [\delta,1]$
    \item $A_{ii}(t) \geq \delta, \forall i$
    \item $A_{ij}(t) > 0 \iff A_{ji}(t) > 0$
\end{enumerate}

For some $\delta > 0$. Assumptions (1) and (2) are what we refer to as regularity assumptions and are relatively mild. They ensure that \textit{all} persistent interactions between agents are edges of the graph $\mathcal{G}_\infty$, and the graph $\mathcal{G}_\infty$ has self-loops\footnote{And is therefore aperiodic.}. Assumption (3) is a bit stronger and ensures that $\mathcal{G}_\infty$ is undirected. Under these Assumptions, it is shown (Lemma 1, \cite{proskurnikov2018tutorial}) that the dynamics will converge, and each connected component of $\mathcal{G}_\infty$ will converge to the same steady state. One of the particularly important consequences of this Lemma is that it implies that even for graphs that are always time-varying (i.e. the sequence of graphs $\{\mathcal{G}(A(t))\}$ do not converge), the \textit{states} of the agents will converge. In other words convergence of states is robust in this instance of time-varying graphs.

Before drawing our conclusions, however, we would like to generalise this a bit further to directed graphs $\mathcal{G}_\infty$, since our private signal models are characterised by directed edges in the form of connections to the ghost nodes. We make use of the following conventions. A strongly connected component (SCC) is a ``sink'' SCC (SSCC) if all outgoing paths from the SCC are reciprocated. Intuitively, these SCCs form the sinks in the meta-graph constructed from SCCs of the original graph. A graph consists of SSCCs if for every path between nodes $i$ and $j$, a path from $j$ to $i$ exists. An obvious and important case of an SSCC-only graphs are strongly connected graphs, but it also includes graphs with multiple fully disconnected strongly connected components. All nodes outside the SSCCs are denoted as ``quasi-connected''. Intuitively, a node $i$ is quasi-connected if it possesses a path to some $j$ and the path $j$ to $i$ does not exist. A graph that contains quasi-connected nodes is a quasi-connected graph. It is easy to see that all nodes are either quasi-connected or a member of a SSCC.

\begin{theorem}\label{thm:scc}
Suppose Assumptions (1) and (2) hold for $t \geq t_0$. All sink strongly connected components of $\mathcal{G}_\infty$ will converge to consensus.
\end{theorem}

Importantly this means that all strongly connected graphs $\mathcal{G}_\infty$ converge to consensus. We can also immediately see that:

\begin{cor}\label{cor:scc}
Suppose Assumptions (1) and (2) hold for $t \geq t_0$. If $\mathcal{G}_\infty$ consists only of SSCCs and without any isolated nodes, $\mathcal{H}(x^*) = 0$ for the steady state $x^*$.
\end{cor}

That is to say, under quite mild regularity conditions, if our averaging dynamics can be represented by a sequence of updates $\{A(t)\}$ where mutually influencing agents exist, heterogeneity is impossible. Put differently, quasi-connectedness in the infinite graph is a necessary condition for heterogeneity. 

Quasi-connected graphs $\mathcal{G}(t)$ are represented by update matrices $A(t) = \bigl( \begin{smallmatrix}Q(t) & R(t)\\ 0 & S(t)\end{smallmatrix}\bigr)$, where $S(t)$ represents edges within SSCCs, $Q(t)$ represents edges between quasi-connected nodes and $R(t)$ represents edges from quasi-connected nodes to SSCCs. Clearly, our private signal  models with $\Lambda > 0$ will fulfil this. Other common classes of models that fulfil these are DeGroot models with \textit{leaders} or \textit{stubborn agents}. Such models assume there exist a set of nodes $\mathcal{L}$ that possess self-weight of $1$, meaning each such agent is an (isolated) SSCC, and all nodes that have a path to such nodes are quasi-connected. It is worth pointing out that when it comes to modelling real-life phenomenon, models with private signals can often be interpreted as equivalent to ones with stubborn agents or leaders. For example, modelling information diffusion on social networks such as Twitter poses a methodological question: do influential accounts such as politicians or celebrities function as news sources or non-reciprocating members of an otherwise peer-to-peer network? In context of these models, both interpretations are likely to be functionally identical.

While quasi-connected (directed) graphs provide us with a necessary condition for heterogeneity, we do unfortunately lose the robust convergence properties we observed for the undirected graphs. We saw that in the undirected case the convergence of the graph was not necessary to ensure convergence of the states of the agents, which is no longer applicable to the quasi-connected case. In fact, we can characterise a general convergence criteria:

\begin{theorem}\label{thm:converge}
Suppose Assumptions (1) and (2) hold for $t \geq t_0$. Let edges between quasi-connected nodes, SSCCs and quasi-connected nodes to SSCCs on $\mathcal{G}_\infty$ be represented in each $A(t)$ by $Q(t)$, $S(t)$ and $R(t)$ respectively. Then the model converges if and only if for all quasi-connected nodes one of the following conditions is met:
\begin{itemize}
    \item $Q(t) \to Q$ and $R(t) \to R$
    \item $R(t)S = (I-Q(t))M + \epsilon(t)$
\end{itemize}
Where $\epsilon(t) \to 0$, $M$ is an arbitrary row-stochastic matrix, and $S = \underset{t\to \infty}{\lim}S(t:t_0)$.
\end{theorem}

The key takeaway from this result is that convergence under quasi-connected graphs is much more \textit{fragile}. Convergence requires that the edges of all quasi-connected nodes converge, or alternatively that edge weights are asymptotically affine transformations of each other. Independent variation in $Q(t)$ and $R(t)$ will generally invalidate the ability of the model to converge.

If the model does converge, the steady state $x^*$ takes a convenient form. Let $x_{QC}(t)$ and $x_{SC}(t)$ denote the states of the quasi-connected agents and the members of the SSCCs (``sink components'', SC, for brevity) respectively, so that $x(t) = [{x_{QC}(t)}^T {x_{SC}(t)}^T]^T$. We can then see:

\begin{cor}\label{cor:ss}
    If the model converges according to Theorem \ref{thm:converge}, then the steady state outcomes are $x^*_{SC} = S x_{SC}(t_0)$ and $x^*_{QC} = M x_{SC}(t_0)$, where $S = \underset{t\to \infty}{\lim}S(t:t_0)$. If the first condition of Theorem $\ref{thm:converge}$ is met, $M = (I-Q)^{-1}R$, otherwise it is the arbitrary row stochastic matrix $M$.
\end{cor}

The corollary implies that the steady state outcomes are purely a function of the states of the SSCC nodes at the time step $t_0$. In many models of interest (for example in stationary models), $t_0 = 0$, which means that the asymptotic outcomes of all agents are determined exclusively by potentially a small minority.

The steady states of the SSCCs will, by Theorem \ref{thm:scc} converge to a consensus in each SSCC, regardless of the structure of $x_{SC}(t_0)$ or $S$, and will therefore have no heterogeneity. On the other hand we can see that the steady states of the quasi-connected nodes have no special structure that encourages such consensus. Let $M_i$ denote the $i$-th row of the matrix $M$. Therefore for all pairs $i$ and $j$ of quasi-connected nodes, $x_i^* = x_j^* \iff (M_i - M_j)\perp x_{SC}(t_0)$. Without any further restrictions on the structure of $M$ or $x_{SC}(t_0)$, this will be violated almost everywhere in the set of possible $M$ and $x_{SC}(t_0)$. We consider technical details in the Appendix.


\subsection{Private signal models}

We now apply these general results to the specific case where we are modelling an augmented graph with ghost nodes representing a model that may incorporate private signals. Let the infinite graph of the augmented interactions be represented by $\tilde{\mathcal{G}}_\infty$ and the infinite graph of the original interactions be represented by $\mathcal{G}_\infty$. Let us assume for simplicity that $\mathcal{G}_\infty$ is strongly connected\footnote{The results can be generalised for quasi-connected graphs where each node is path connected in $\mathcal{G}_\infty$ to at least one node $i$ where $\lambda_i > 0$. We discuss this briefly in the Appendix.}. In this case, we can see that $Q(t) = (I-\Lambda)A(t)$, $R(t) = \Lambda W(t)$ and $S(t) = I$. That is, the SSCCs are just singleton ghost nodes, and \textit{all} the original nodes are quasi-connected.

From this representation we can quickly conclude from Theorem \ref{thm:scc} that for \textit{any} model that can be represented by a strongly connected interaction graph (and with regularity assumptions fulfilled), \textit{private signals are a necessary condition to model heterogeneity}. For example, a very common set of assumptions in models of dynamics over graphs is that the graph (and adjacency matrix $A$) is stationary and strongly connected. We can see that any attempt to model averaging dynamics over this graph will require incorporating private signals for us to observe the rich heterogeneous outcomes that characterise real processes.

We can translate Theorem \ref{thm:converge} into the following Corollary:

\begin{cor}\label{cor:convergeprivate}
Consider a quasi-linear updating model where $\Lambda \neq 0$ and $\mathcal{G}_\infty$ is strongly connected. Suppose assumptions (1) and (2) hold for $t \geq t_0$. The model converges if and only if one of the following conditions are met:
\begin{itemize}
    \item $A(t) \to A$ and $W(t)C = b(t) \to b$
    \item $\Lambda W(t) = (I-(I-\Lambda)A(t))M + \epsilon(t)$
\end{itemize}
Where $\epsilon(t) \to 0$ and $M$ is an arbitrary row-stochastic matrix.
\end{cor}

Here again we see that the conditions for convergence for a model with private signals are much stricter. Either both the interaction matrix $A(t)$ \textit{and} the private signals $b(t)$ must converge, or again are asymptotically affine transformations of one another. 

With the structure of private signal models, we can go further and refine the conditions to observe heterogeneity in the steady state. It is straightforward to show that:

\begin{cor}\label{cor:hetero}
Consider a quasi-linear updating model where $\Lambda \neq 0$ and $\mathcal{G}_\infty$ is strongly connected. Suppose assumptions (1) and (2) hold for $t \geq t_0$. If the model converges, then the steady state is heterogeneous only if both $A(t)$ and $b(t)$ converge, or neither of them do.
\end{cor}

In other words, if the convergence of the two elements $A(t)$ and $b(t)$ are mismatched, then heterogeneity will not be achieved. As we show in the Appendix, this follows from the fact that for a convergent model, if $A(t) \to A$, then $b(t) \to b$, and if $b(t) \to b$, $A(t)$ can keep oscillating only if the steady state is fragmented or at consensus.

This result can be quite useful if we know for example that one of the update matrices $A(t)$ or $b(t)$ is a function of $x(t)$, in which case any convergence of the latter will guarantee convergence of the update matrices. For example, suppose that $b(t) = b(x(t))$. In this case for any convergent model ($x(t) \to x^*$), we can conclude that $b(t) \to b(x^*) = b^*$. Therefore if $A(t)$ continues to vary independently, we know that $x^*$ must be at consensus or fragmented. A typical case in which this occurs is if the neighbourhood updates are asynchronous (i.e. a subset of vertices update their states as a function of their neighbours). In such models, without any need to evaluate the dynamics, we know that heterogeneity will never be achieved. We consider an example of such in Section \ref{sec:swarm}.

\subsubsection{Steady state outcomes}

We are generally more interested in the first condition for convergence in Corollary \ref{cor:convergeprivate} (as the latter is mostly used to illustrate the restrictiveness of the conditions that allow for convergence - note that the second condition implies the first). If the model converges under this condition, we can see that the steady state of the original nodes will be:

\begin{equation}\label{eq:ss}
    x^* = (I - (I-\Lambda)A)^{-1} \Lambda W C = \underbrace{(I - (I-\Lambda)A)^{-1}}_{F}\underbrace{\Lambda b}_{B} = F B
\end{equation}

Here $B \in \mathbb{R}^{N \times d}$ is a (weighted) vector of the asymptotic private signals that agents receive. The matrix $F \in \mathbb{R}^{N \times N}$ is denoted as the fundamental matrix in reference to the analogous object in the theory of absorbing Markov Chains \cite{levin2017markov}.

There are a few interesting implications of the steady state expression in Equation \ref{eq:ss}. If the model is exogenous (i.e. $A(t)$ and $b(t)$ evolve independently of the state $x(t)$, or are stationary), we can see that the early stage dynamics of the process have \textit{no} impact on the steady state outcome. That is, since $x^*$ is purely a function of the asymptotic $A$ and $b$ we can effectively ignore all intermediate states when we determine the steady state. For example, suppose that $A(t)$ and $b(t)$ are realised stochastically, but we can show independently that the noisy realisations converge almost surely to $A$ and $b$. In this case only the limits are required. In this sense exogenous models with private signals are \textit{ergodic} - the initial state $x(0)$ has no impact on the steady state outcomes\footnote{Apart from, for example, very specific modelling choices such as $b = x(0)$ as in the original Friedkin-Johnsen model \cite{friedkin1990social}.}.

Of course, this restriction does not necessarily apply to endogenous models (where $A(t) = A(x(t))$ or $b(t) = b(x(t))$). We can see therefore that if we are attempting to model phenomenon that demonstrate both hysteresis and heterogeneity (with regularity assumptions), endogeneity will be a necessary element to incorporate into our averaging dynamics.

\subsubsection{Information diffusion and random walks}\label{ss:diffusion}

Equation \ref{eq:ss} provides a useful decoupling of the role of topology (summarised by $F$) and the distribution of private signals (summarised by $B$) on the steady state outcomes of a model with private signals. Furthermore, the form of the fundamental matrix provides a useful conceptual bridge to the theory of random walks and information diffusion over graphs. In this section we briefly introduce some of these concepts to show how they can help us intuitively interpret the equilibrium outcomes of many multi-agent system models.

Consider the augmented form of the dynamics as represented in the block matrix in Equation \ref{eq:block}, except we consider the asymptotic form:

\begin{equation}
      \underset{t \to \infty}{\lim} \tilde{A}(t) = \tilde{A} = \left[
        \begin{array}{c | c}
        (I-\Lambda) A & \Lambda W\\
        \hline
        0 & I
        \end{array}
        \right]
\end{equation}

We can see this can be interpreted as the transition matrix of an absorbing Markov Chain, where the ghost nodes represent absorbing states and the original nodes represent transient states. The limiting powers of this matrix therefore encode the probability of a random walk starting at any of the transient states and ending up at any of the absorbing states:

\begin{equation}
      \underset{t \to \infty}{\lim} \tilde{A}^t = \left[
        \begin{array}{c | c}
        0 & (I - (I-\Lambda)A)^{-1} \Lambda W\\
        \hline
        0 & I
        \end{array}
        \right] = 
        \left[
        \begin{array}{c | c}
        0 & F \Lambda W\\
        \hline
        0 & I
        \end{array}
        \right]
\end{equation}

In the language of Markov Chains, the fundamental matrix $F$ encodes in element $(i,j)$ the expected number of times a walk that begins at node $i$ at time $0$ hits node $j$ over its lifetime. At some finite time, all random walks must ``exit'' the set of transient states because they will hit an absorbing state and remain fixed there. The probability of this happening at any transient state $i$ is $\lambda_i$. The product $\Lambda W$ therefore encode for each transient set the probability of moving to each possible absorbing state. The $(i,g)$-th entry of the product $F \Lambda W$ encodes for each node $i$ the total probability that a random walk beginning at $i$ ends up being absorbed at absorbing state $g$.

The steady state of our model is $x^* = F\Lambda W C = F B$. Therefore, we can conclude that the steady state of a specific node $i$, $x_i^*$ can therefore be seen to be a weighted average between all ghost nodes, with the weight for ghost $g$ being the probability of a random walk starting at $i$ hitting node $g$ first under the asymptotic dynamics $A$. For the sake of illustration suppose $\mathcal{X} = [0,1]^N$. That is, the agent states are one-dimensional and we can represent private signals as the weight between two ghost nodes, one with state $0$ and one with state $1$. The steady state outcome for each node $i$ just encodes the probability of hitting the positive ghost node first.

An alternative but useful re-characterisation of this same process considers the diffusion of information \textit{from} the private signals \textit{to} the original ghost nodes. Suppose for illustration again that the states of the agents are one dimensional with $x_i \in [-1,+1]$. Consider the following model: each agent $i$ accrues an information set $\mathcal{I}_i(t)$ of discrete signals where each signal takes value $+1$ or $-1$. At each time step, the agent $i$ obtains a single new signal $s_i(t)$ in one of two ways. A signal is drawn from a neighbour $j$'s previous signal $s_j(t-1)$ with probability $(1-\lambda_i) A_{ij}$. Alternatively with probability $\lambda_i$ the agent draws a signal from a ghost node. The positive ghost node $g_+$ produces signals of only $+1$ the negative ghost node $g_-$ produces signals of only $-1$. The ghost nodes are drawn with probabilities $\lambda_i W_{i+}$ and $\lambda_i W_{i-}$ such that $W_{i+} - W_{i-} = b_i$.

\begin{figure}
    \centering
    \includegraphics[width=\textwidth]{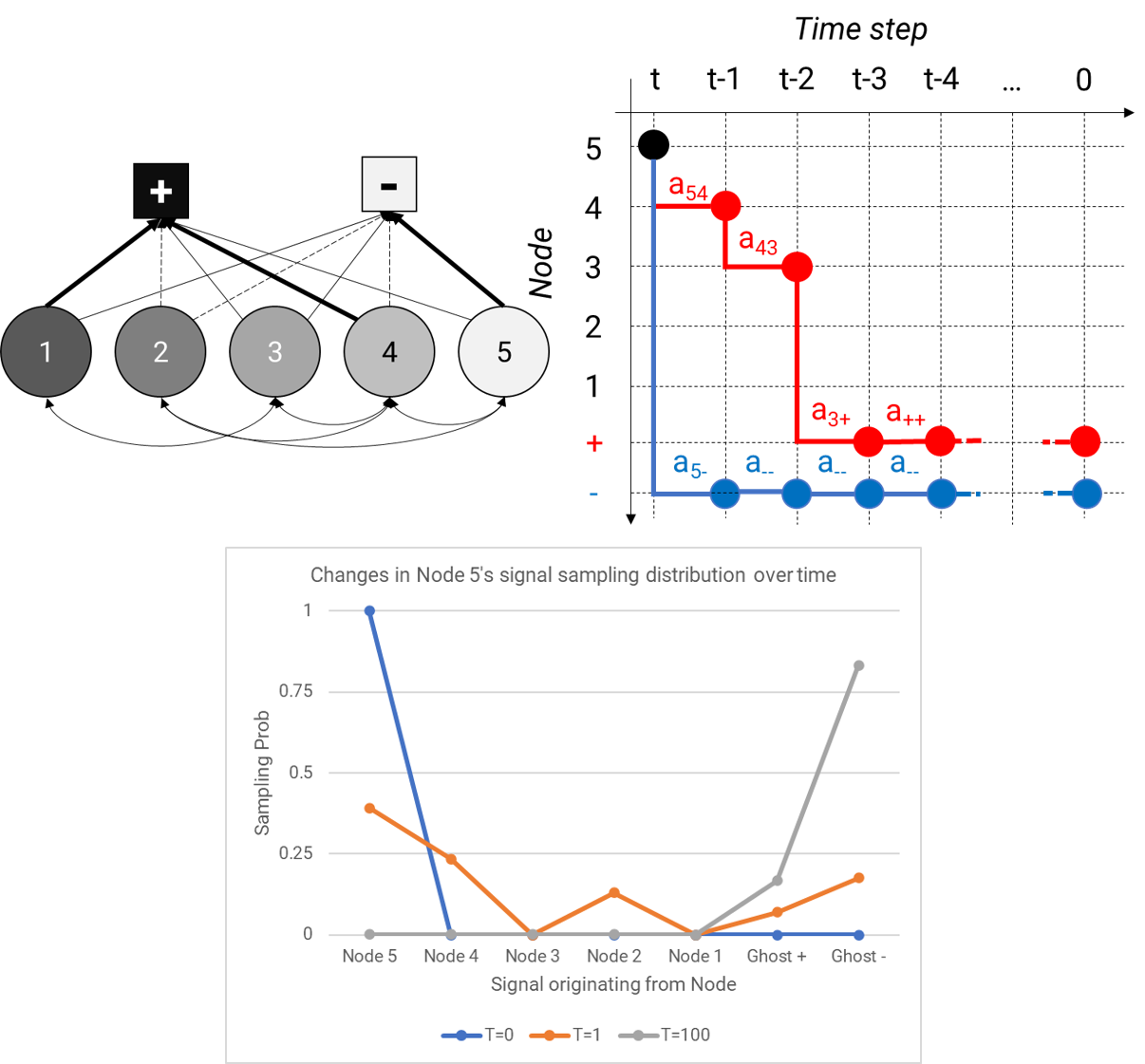}
    \caption{{Illustrating the ideas of ``contact tracing''. For the graph $\mathcal{G}$ in the top left, we consider tracing the potential pathways of signals that appear in node 5's information set at time $t$, with two potential paths illustrated. We can use this to construct the sampling distribution of signals drawn by node $5$ for various times $t$. We demonstrate this for $t=0$ where it is concentrated on its own original signal, and the distribution converges over time to the limiting probability of the ghost nodes.}}
    \label{fig:contact}
\end{figure}

We are interested in establishing the asymptotic composition of the information set $\mathcal{I}_i(t)$. In order to do so consider the probability that a signal $s_i(t)$ drawn from some large $t$ is positive ($\mathbb{P}[s_i(t)=+1]$). This is simply the probability that the signal originated from the positive ghost node $g_+$\footnote{Since $t$ is large, the probability that a signal arrives from a ghost node approaches $1$.}. For example, the agent $i$ could have sampled from neighbour $j$ at $t$, and the neighbour $t$ sampled from the positive ghost node at $t-1$. Each of these possible pathways ($i \to j \to \ldots \to g_+$) is equivalent to a random walk from $i$ to $g_+$. We use the analogy of ``contact tracing'' to describe these possible pathways, and illustrate the general idea in Figure \ref{fig:contact}, which shows how the probabilities $\mathbb{P}[s_5(t)=+1]$ and $\mathbb{P}[s_5(t)=-1]$ converges over time to a fixed distribution for draws at that agent. This fixed distribution therefore establishes the asymptotic composition of the accrued information set $\mathcal{I}_i(t)$.

In sum, we can see that the steady state outcomes of the agents in a model of averaging dynamics with private signals can be conceptualised as the aggregation of signals that are transmitted from a small set of ghost nodes. Of course, this is just a characterisation of the steady state (asymptotic) outcome: it is by no means precisely what is happening in the intermediate dynamics. Recall for example that the intermediate dynamics can have varying $A(t)$ or $b(t)$, which can have no relation to the random walks characterised by the asymptotic $A$ and $W$.

Nonetheless, this random walk interpretation can be useful to build intuition about the steady state outcomes we might expect for different agents, so long as we have some sense of the asymptotic graph $\mathcal{G}(A)$. For example, if the clustering coefficient of $\mathcal{G}(A)$ is high\footnote{For example, if the graph is constructed to connect $k$-nearest neighbours over some metric space, there is likely to be high transitivity.}, it means that random walks that begin at any node $i$ will circulate with high probability back to $i$. Therefore, each node will place a larger weight on their own private signals than in a comparative graph with a low clustering coefficient. If nearby agents receive similar private signals (a form of ``homophily'') we can see that the signals that get circulated locally tend to only be of a single type, so the steady state outcome of neighbouring nodes will be highly correlated, in contrast to a graph where private signals are independently distributed amongst all agents. If private signals are homophilous, but there exist hubs in the network, we can see that signals can travel very ``far'' in the network, and local correlations are mitigated in the steady state outcomes.

\section{Examples and Applications}

In this section we consider an intentionally diverse set of models of social, economic and machine behaviour to illustrate how the frameworks we discussed can be applied.

\subsection{Contrarian agents}

\begin{figure}
        \centering
        \begin{subfigure}[b]{0.475\textwidth}
            \centering
            \includegraphics[width=\textwidth]{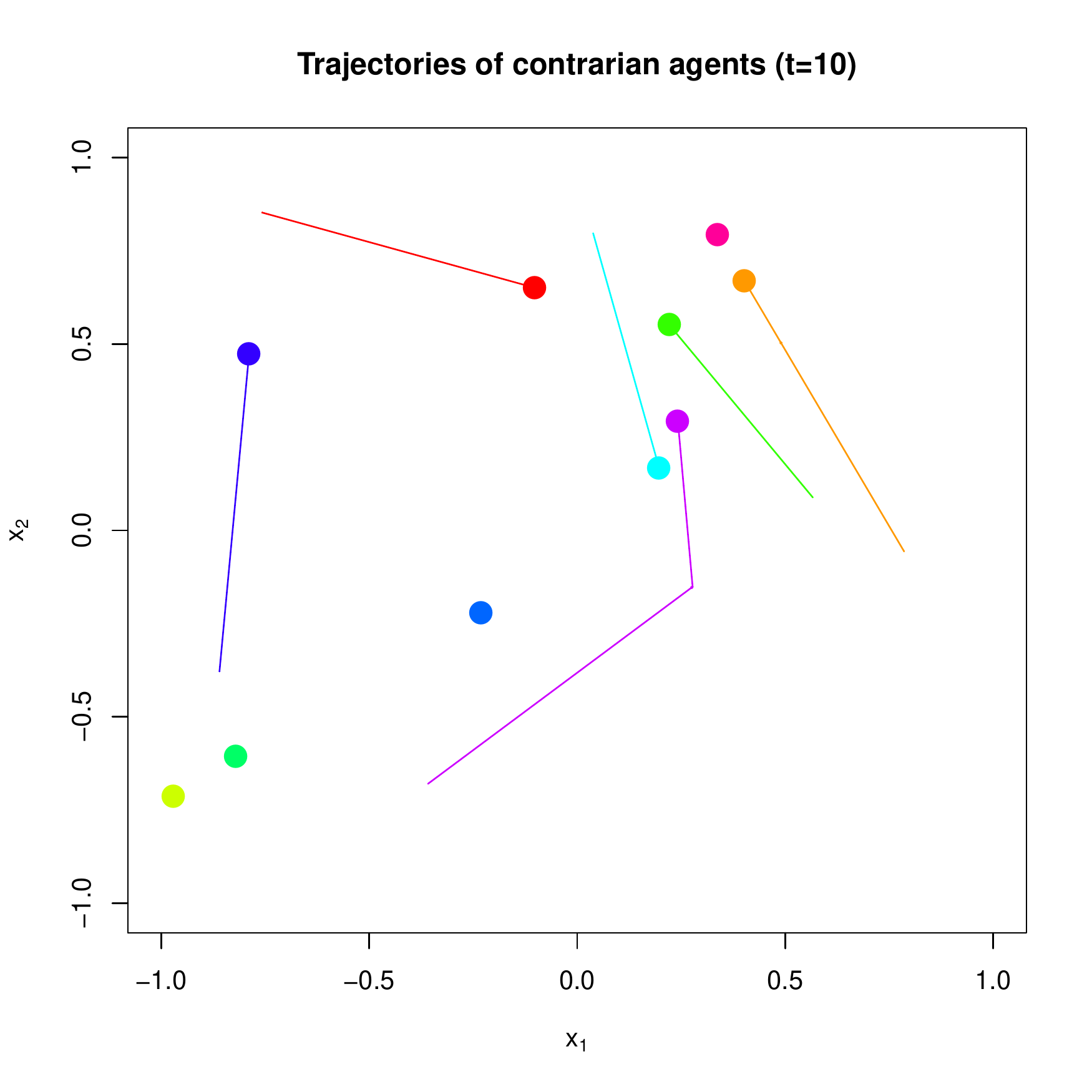}
            \caption[Contrarian agents at $t=10$]%
            {{\small Contrarian agents at $t=10$}}
        \end{subfigure}
        \hfill
        \begin{subfigure}[b]{0.475\textwidth}  
            \centering
            \includegraphics[width=\textwidth]{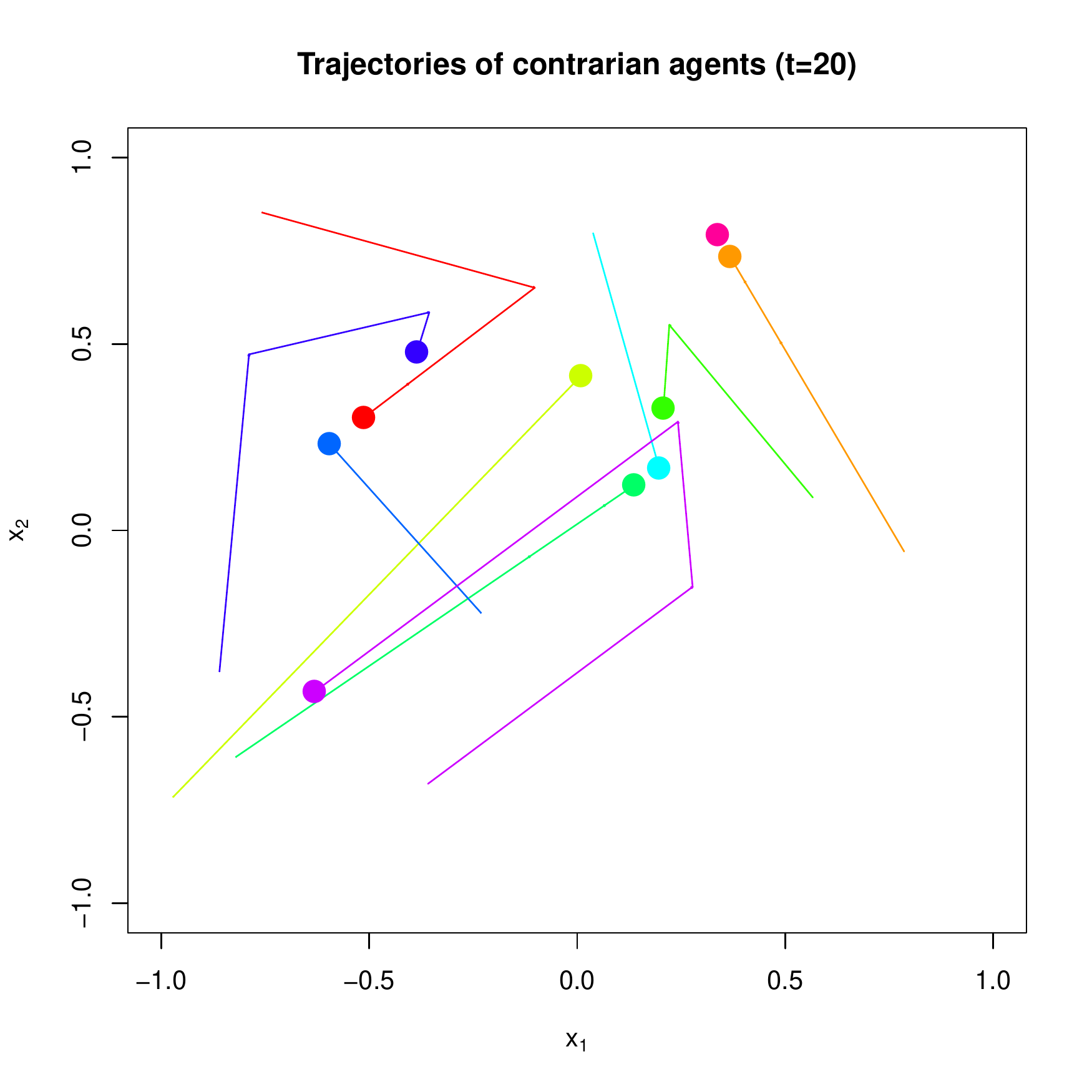}
            \caption[Contrarian agents at $t=20$]%
            {{\small Contrarian agents at $t=20$}}
        \end{subfigure}
        \vskip\baselineskip
        \begin{subfigure}[b]{0.475\textwidth}   
            \centering
            \includegraphics[width=\textwidth]{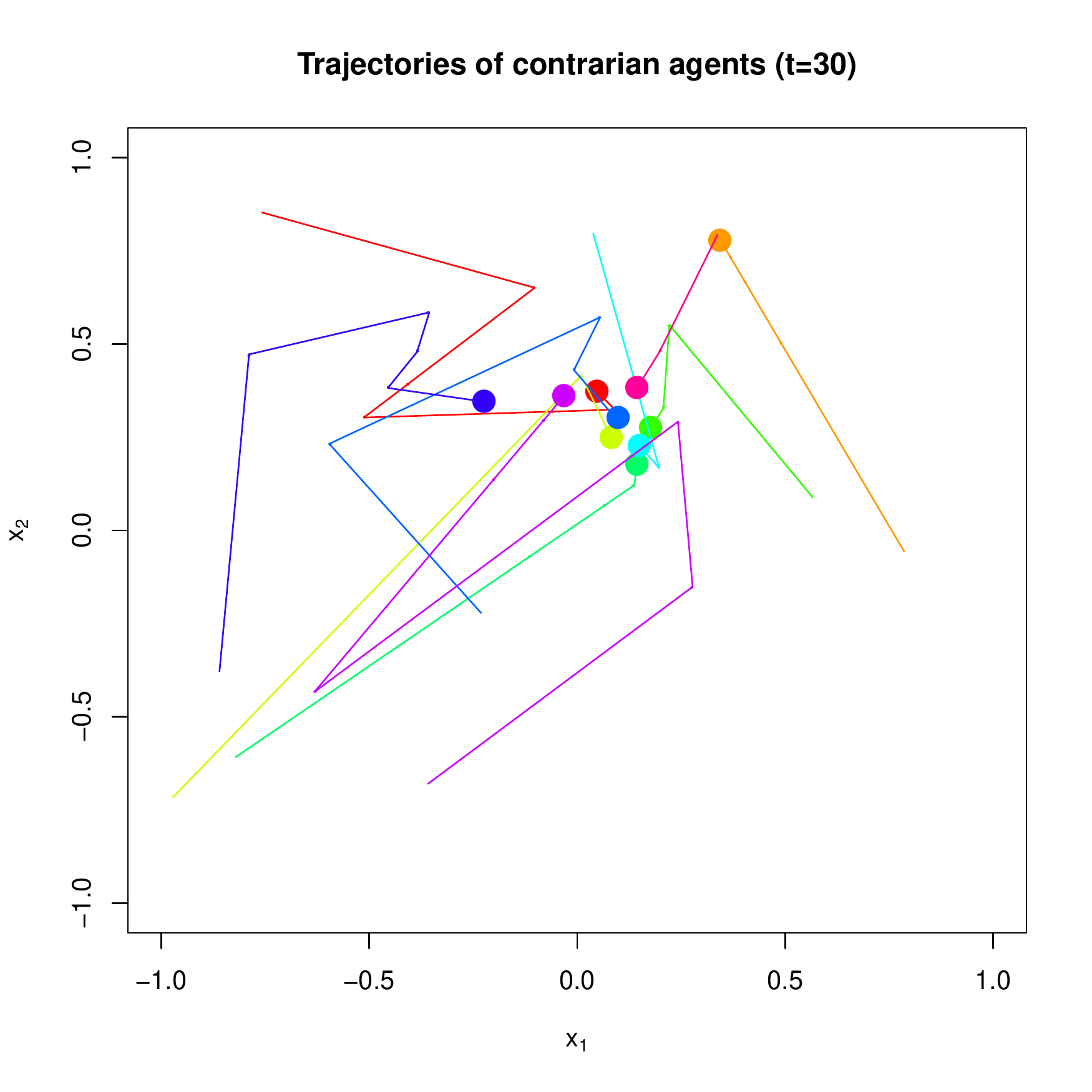}
            \caption[Contrarian agents at $t=30$]%
            {{\small Contrarian agents at $t=30$}}
        \end{subfigure}
        \quad
        \begin{subfigure}[b]{0.475\textwidth}   
            \centering 
            \centering
            \includegraphics[width=\textwidth]{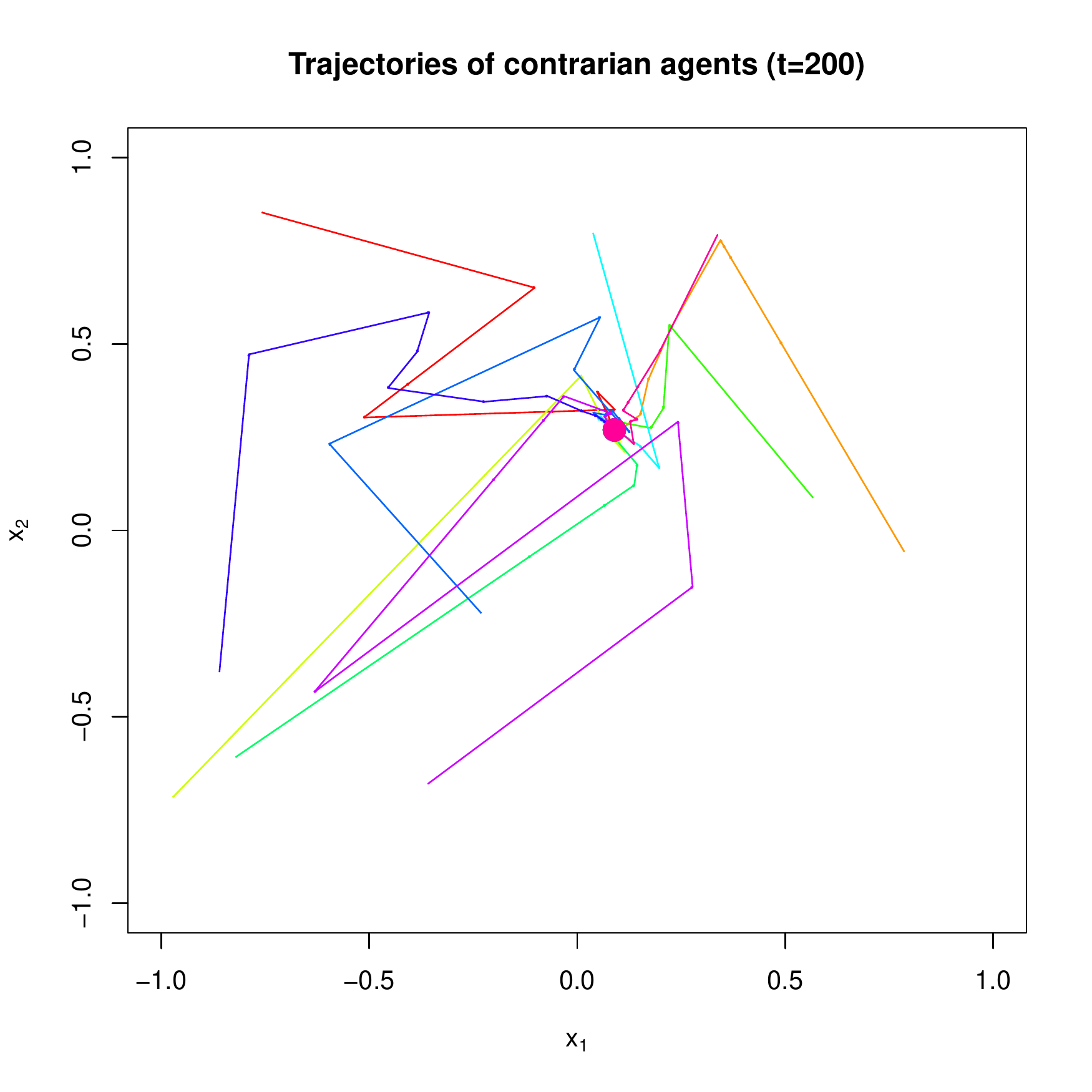}
            \caption[Contrarian agents at $t=200$]%
            {{\small Contrarian agents at $t=200$}}
        \end{subfigure}
        \caption[ The two-dimensional trajectories of contrarian agents towards consensus. ]
        {\small The two-dimensional trajectories of contrarian agents converging towards consensus. The current position of an agent is denoted with a large circle and their trail of past positions is a solid line with the same colour. In the bottom right panel we can see the states have converged.}
        \label{fig:contrarian}
    \end{figure}
    
Suppose there exist a set of $N$ agents that are connected over some latent strongly connected graph $\mathcal{G}$. Each agent possesses a state $x_i(t) \in X \subset \mathbb{R}^d$. At every time step $t$, a randomly chosen subset of agents will update their states (i.e. updating is asynchronous).

Agents in this model are contrarians, and they prefer to update their state towards observed states that are as different as possible from their current states, and ignoring neighbours with similar states. We can think of this as modelling for example, agents with multi-dimensional opinions, and who are most influenced by friends with ``surprising'' opinions. Alternatively, we can also consider the agents as representing financial actors and $x_i(t)$ some representation of their investment strategy. If they observe neighbours that take very different strategies, they may assume that neighbour possesses private information, and switch to mimic the behaviour.

In order to pick discordant neighbours, each updating agent will observe the states $x_j(t)$ of her neighbours $j \in \mathcal{N}(i)$ over the underlying graph $\mathcal{G}$ and measure the distance to her own state according to some metric $\norm{.}$ over $X$. A neighbour $j$ is picked with probability:

\begin{equation}
    p_{ij} = \frac{\norm{x_i(t) - x_j(t)}}{\sum_{k \in \mathcal{N}(i)} \norm{x_i(t) - x_k(t)}}
\end{equation}

The agent $i$ then updates their state towards the chosen neighbour $j$:

\begin{equation*}
    x_i(t+1) = \gamma x_i(t) + (1-\gamma) x_j(t)
\end{equation*}

The parameter $\gamma > 0$ just modulates the speed of updating. If it is high, then agents update their states slowly, and vice versa if it is low. We can see that the update matrix $A(t) = A(x(t))$ varies at each time step as only a subset of edges are activated. The updates are not symmetric (i.e. Assumption 3 used in the convergence of undirected infinite graphs is violated). Nonetheless, the infinite graph will still be strongly connected, and we can see through the parameter $\gamma > 0$ that realised edges will not decay to $0$, fulfilling our regularity assumptions. Therefore, conditions for Theorem $1$ are met, and we can conclude that the dynamics will converge to a consensus.

We illustrate this in Figure \ref{fig:contrarian}, where we take $X = [-1,1]^2$, $N=10$, $\gamma = 0.1$ over a strongly connected directed Erdos-Renyi graph, and use the Euclidean norm to measure distance between nodes. We can see that despite the early dynamics of the agents being somewhat haphazard as they try to gravitate away from nearby nodes, the dynamics eventually converge.

\subsection{Swarm behaviour}\label{sec:swarm}

\begin{figure}
        \centering
        \begin{subfigure}[b]{0.475\textwidth}
            \centering
            \includegraphics[width=\textwidth]{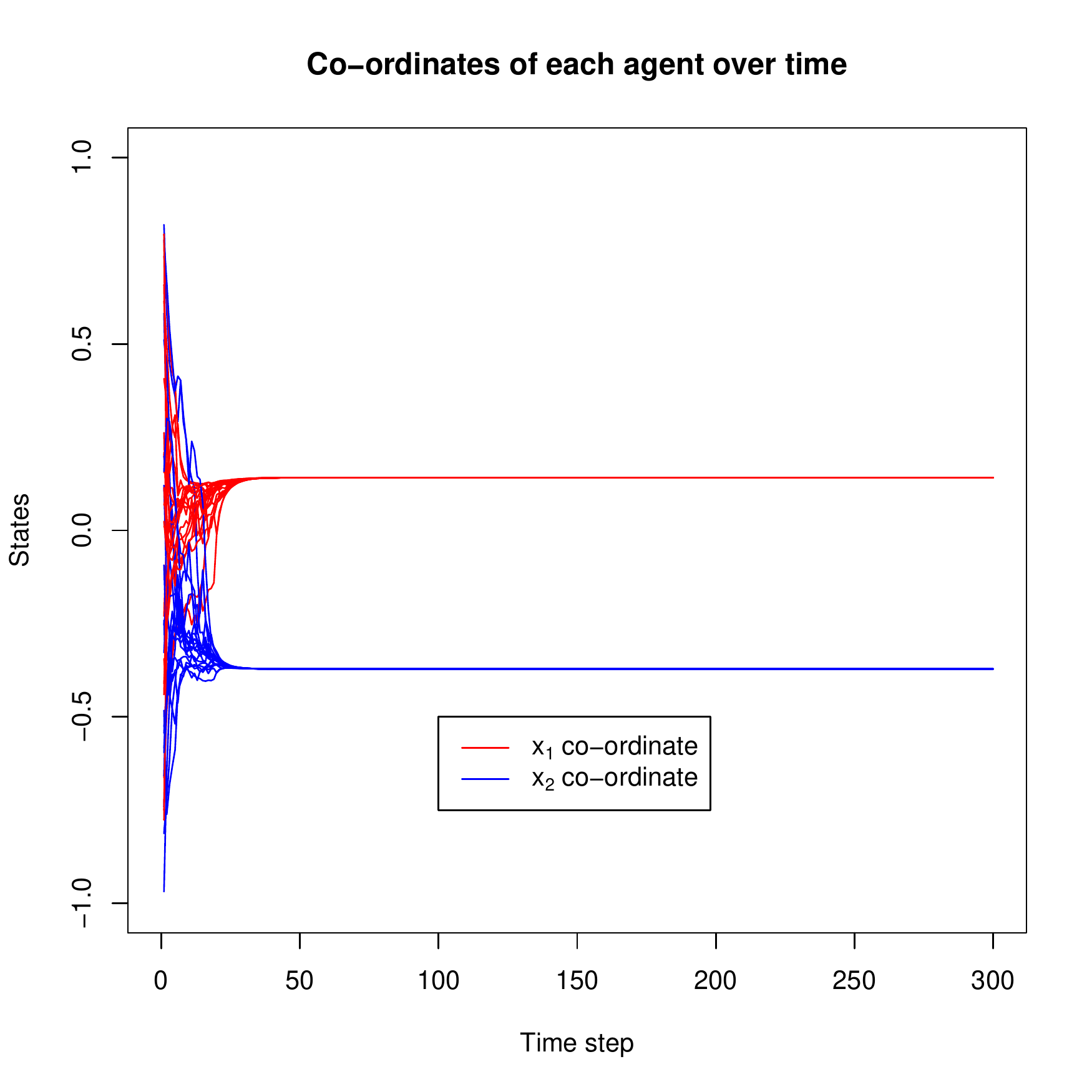}
            \caption[Swarm $x_1$ and $x_2$ co-ordinates over time (asynchronous).]%
            {{\small Swarm $x_1$ and $x_2$ co-ordinates over time (asynchronous).}}
        \end{subfigure}
        \hfill
        \begin{subfigure}[b]{0.475\textwidth}  
            \centering
            \includegraphics[width=\textwidth]{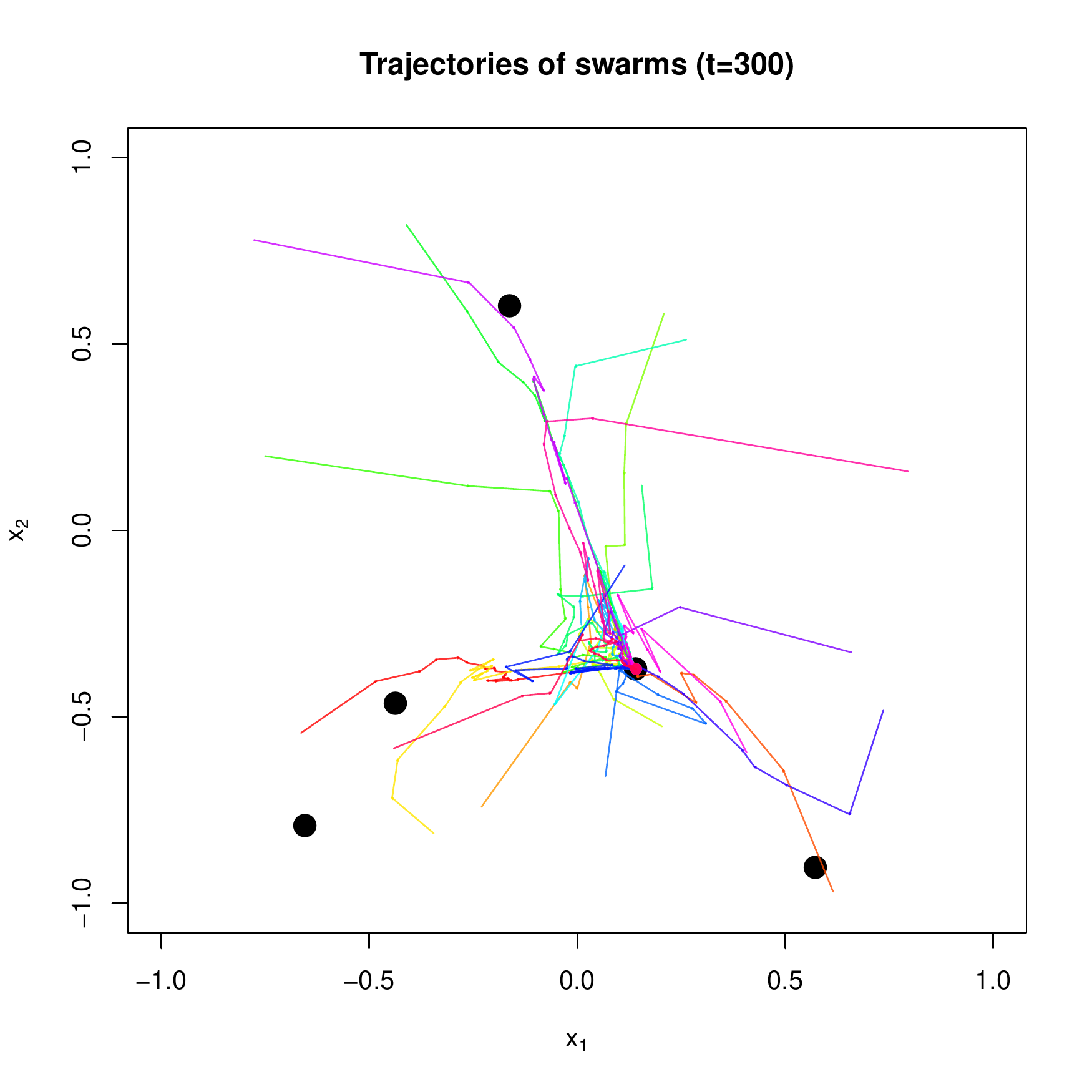}
            \caption[Swarm trajectories at $t=300$ (asynchronous).]%
            {{\small Swarm trajectories at $t=300$ (asynchronous).}}
        \end{subfigure}
        \vskip\baselineskip
        \begin{subfigure}[b]{0.475\textwidth}   
            \centering
            \includegraphics[width=\textwidth]{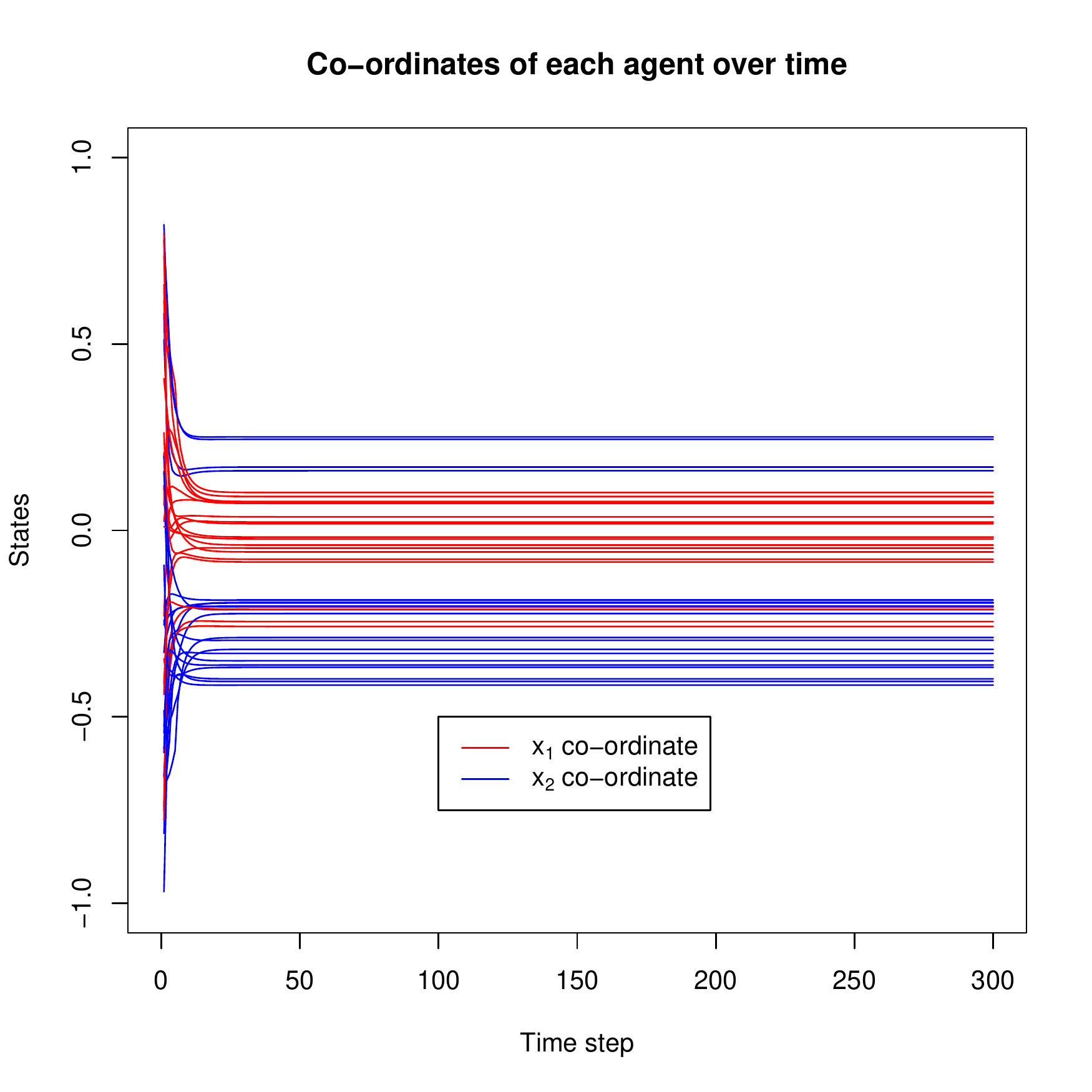}
            \caption[Swarm $x_1$ and $x_2$ co-ordinates over time (synchronous).]%
            {{\small Swarm $x_1$ and $x_2$ co-ordinates over time (synchronous).}}
        \end{subfigure}
        \quad
        \begin{subfigure}[b]{0.475\textwidth}   
            \centering 
            \centering
            \includegraphics[width=\textwidth]{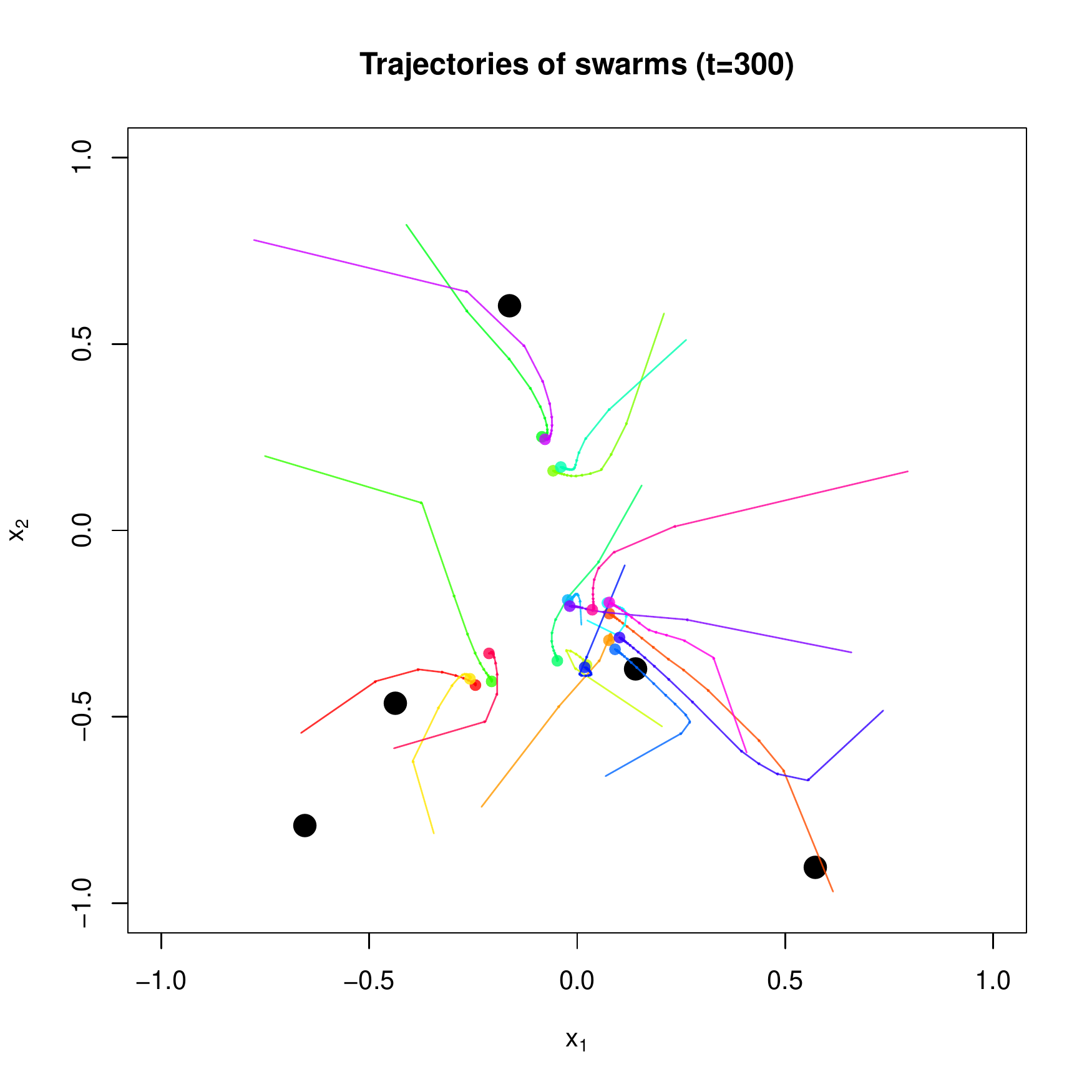}
            \caption[Swarm trajectories at $t=300$]%
            {{\small Swarm trajectories at $t=300$ (synchronous).}}
        \end{subfigure}
        \caption[ The two-dimensional trajectories of contrarian agents towards consensus. ]
        {\small The two-dimensional trajectories of swarms with asynchronous updating (top panels) and synchronous updating (bottom panels). The left hand panels denote the $x_1$ and $x_2$ co-ordinates of all the agents over time. The right hand panels show the 2D trajectories. The current position of an agent is denoted with a large circle and their trail of past positions is a solid line with the same colour. The positions of the landmarks are denoted with solid black circles. In the top right panel we can see the agents have converged to a single point, whereas in the bottom right panel they have all achieved a unique outcome. Both dynamics are realised over the same $k$-regular graph with $N=20$ and $k=3$, and the same starting states uniformly distributed over the unit square. The only difference otherwise is that asynchronous agents randomly sample a subset of neighbours when they update.}
        \label{fig:swarms}
    \end{figure}

In the previous example we showed how strongly connected infinite graphs without private signals converge to a consensus and fail to produce heterogeneous outcomes. We now show that even if private signals are present, heterogeneity may not be achieved if the conditions of Corollary \ref{cor:hetero} are not met. In particular we demonstrate that a lack of co-ordination between the convergence $A(t)$ and $b(t)$ can eliminate or enable heterogeneity in steady state outcomes.

Suppose we have a ``swarm'' of agents located in $[-1,1]^2$ that are attempting to search for food sources (``landmarks'') while trying to not stray too far from a set of preferred neighbours. To model this suppose we have the preferred set of neighbours encoded in an undirected $k$-regular graph (that is, each node has $k$ other agents they are trying to stay nearby\footnote{The choice of graph structure is arbitrary; the argument could equally be made for a fully connected graph (agents have no preferences over neighbours), a hub and spoke graph (i.e. leader/follower structure).}). There are also a set $\mathcal{L} = \{L_1, L_2, \ldots, L_k\}$ landmarks randomly distributed over $[-1,1]^2$. The swarm agents search for landmarks conservatively, moving closer to their closest landmark ($l_i(t)$) at each timestep, but also making sure they do not stray too fair from their neighbourhood. Intuitively, we could suppose that agents get more nutrients the closer they are to a landmark, but do not want to stray too far from their neighbours. That is, each agent updates their position as:

\begin{align}
    x_i(t+1) = \gamma l_i(t) + \frac{(1-\gamma)}{2} x_i(t) + \frac{(1-\gamma)}{2} k_i^{-1}\sum_{j \in \mathcal{N}(i)} x_j(t)
\end{align}

We refer to these as the synchronous dynamics. We can see that Assumptions (1) and (2) are fulfilled, $\mathcal{G}_\infty$ is strongly connected, and $A(t) = A$ is fixed. Furthermore, we can see that if this model converges, then the private signals (the location of the closest landmark), which vary only as a function of $x(t)$, will also converge. As such, this model converges (as per Theorem \ref{thm:converge}) and displays the necessary conditions to achieve heterogeneous outcomes.

On the other hand, consider the following marginal difference: agents choose a random subset of their neighbourhoods at any time step when they update:

\begin{align}
    x_i(t+1) = \gamma l_i(t) + \frac{(1-\gamma)}{2} x_i(t) + \frac{(1-\gamma)}{2} k_i(t)^{-1}\sum_{j \in \mathcal{N}(i,t)} x_j(t)
\end{align}

We refer to these as the asynchronous updates. The only difference is the set of neighbours $\mathcal{N}(i,t)$ (and the degree $k_i(t)$) is now a function of $t$. We can now see that for a convergent model, $b(x(t)) \to b$, but $A(t)$ will vary endlessly, meaning that no heterogeneous steady states can be achieved as per Corollary \ref{cor:hetero}.

We illustrate two representative trajectories in Figure \ref{fig:swarms}, with asynchronous dynamics on the top panels and synchronous dynamics on the bottom panels. We chose $N=20$ and $k=3$, randomly picking locations of $5$ landmarks over the state space. The starting positions of the agents are uniformly drawn over the state space but identical between the two sets of trajectories. This means the only difference between the dynamics is that the synchronous dynamics cause the agent to select $2$ neighbours at each time step and in the asynchronous dynamics $1$ random neighbour is chosen half the time and the other half both neighbours are chosen. We can see despite this minute difference, the trajectories are very different, with the asynchronous dynamics resulting in a consensus as predicted and the synchronous dynamics resulting in a heterogeneous steady state with each agent converging to a unique position.

\subsection{Recommender systems and feedback effects}

We now consider an example where private signals exist explicitly and show how both heterogeneous and consensus outcomes can arise in different parts of the parameter space. We consider an example of machine behaviour with a simple model of how recommender systems might adapt to the preferences of users, a version of which was considered in \cite{sikder2020minimalistic}. Consider a set of $N$ agents over a fixed social network $\mathcal{G}$. Agents possess some state $x_i(t) \in [-1,1]$ that represents their current tastes. Agents update their tastes by interpolating between the tastes of their neighbours and that of a personalised recommender that attempts to provide a signal $\sigma_i(t) = \{-1, +1\}$ that is as close as possible to the current state of its user $x_i(t)$. Clearly, $\sigma_i(t) = \text{sign}(x_i(t))$.

For example, $x_i(t)$ could represent a user's political stance and the recommender engine offers news articles that match an agent's stance. Alternatively, $x_i(t)$ could represent a user's purchase history that favours competing brands ($+1$ or $-1$) and the recommender system offers products that complement a user's past purchases. A typical example of this are technology ecosystems that confer network effects, such as phones (Apple vs Android) and the associated accessories.

\begin{figure}
    \centering
        \begin{subfigure}[b]{0.475\textwidth}
            \centering
            \includegraphics[width=\textwidth]{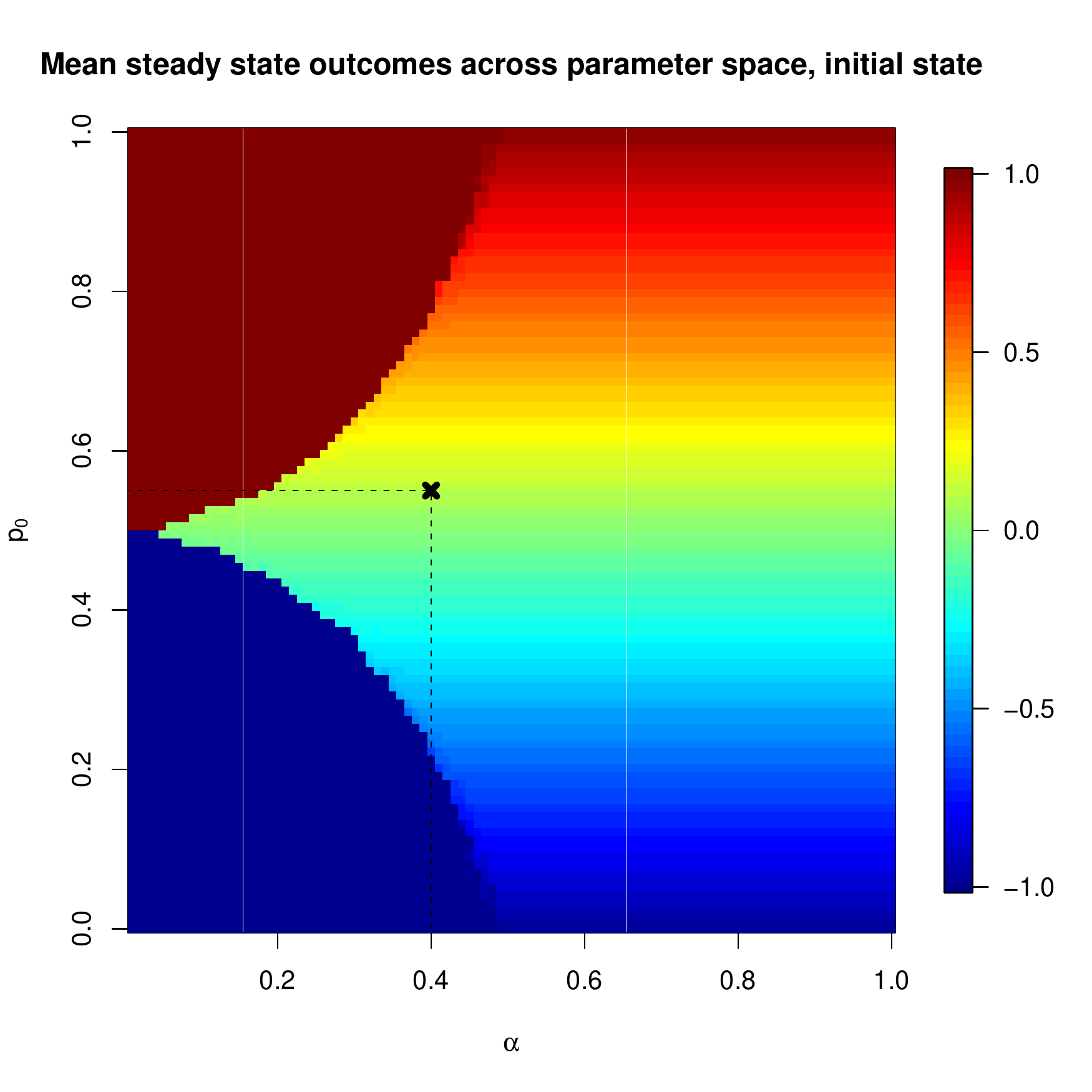}
            \caption[Curtain plot.]%
            {{\small Parameter space demonstrating areas of consensus (dark red/blue) and heterogeneous outcomes.}}
        \end{subfigure}
        \hfill
        \begin{subfigure}[b]{0.475\textwidth}  
            \centering
            \includegraphics[width=\textwidth]{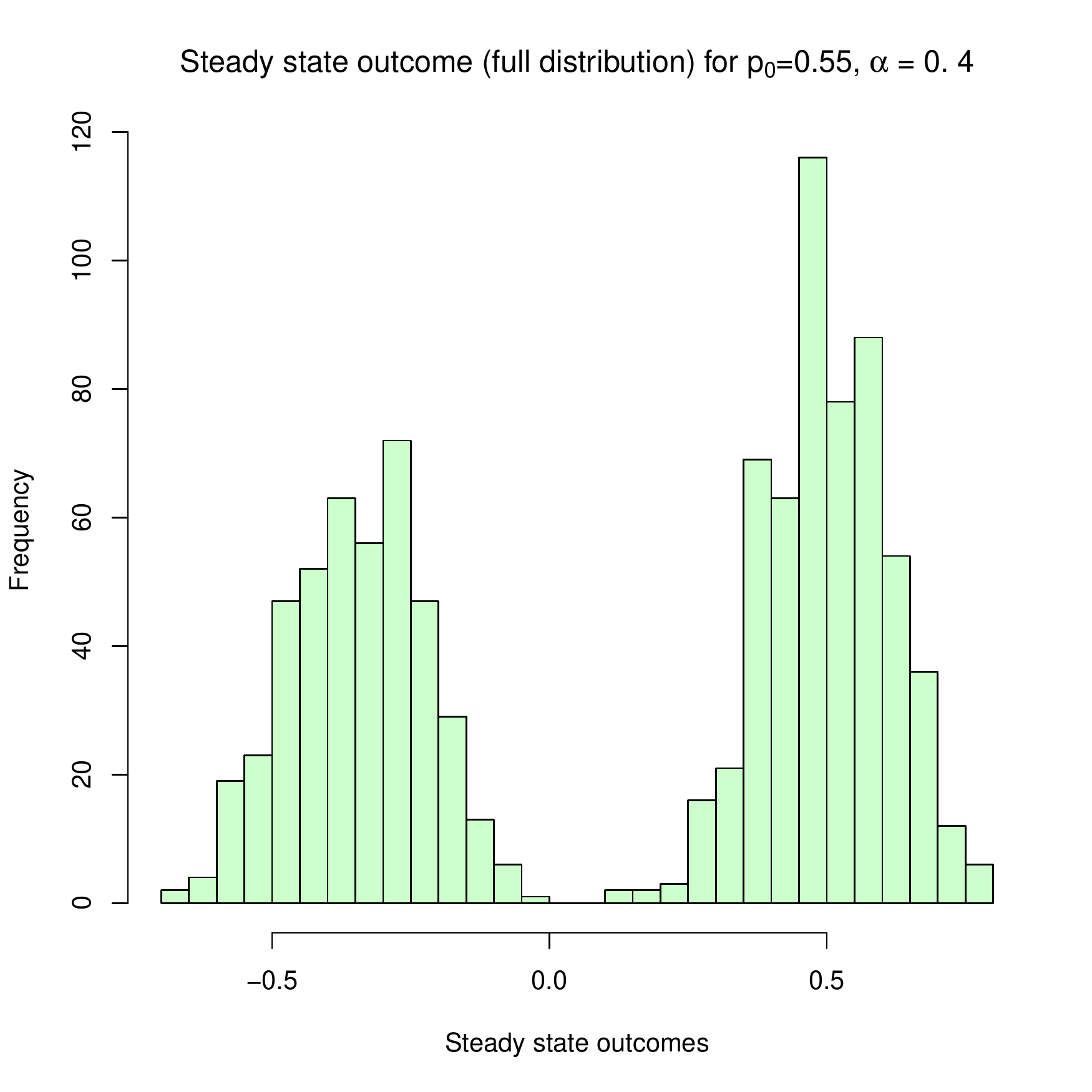}
            \caption[Heterogeneous steady state.]%
            {{\small Example of a heterogeneous steady state.}}
        \end{subfigure}

    
    \caption{A) Demonstration of how a model with private dynamics can result in both heterogeneous and consensus outcomes. The x-axis measures the weight on private signals $\alpha$ and the y-axis is the fraction of positive signals $p_0$ in $\sigma(x(0))$. The gradient at each point is the mean value of steady state outcomes $x^*$, given that the dynamics unfold as per Equation \ref{eq:curtain}. In the top left and bottom left, where $\alpha$ is low and the initial distribution $x(0)$ is weighted towards positive or negative signals, the dynamics cascade so that all agents (user and recommender) converge to either $+1$ or $-1$. The region in between denotes outcomes where the recommenders do not cascade, the steady state mean outcome is not at an extreme, and heterogeneous outcomes are supported. B) An example of a heterogeneous outcome, where $p_0 = 0.55$, $\alpha=0.4$ (as indicated by a cross on the left panel). Numerical simulations were conducted over an Erdos-Renyi graph with $\langle k \rangle  = 12$, $N=1000$, and where $a_{ij} = \frac{1}{k_i}$ if $i,j$ were connected on $\mathcal{G}$.}
    \label{fig:curtain}
\end{figure}

If we are not interested in modelling social network effects, the dynamics can also be extended to understand how correlated tastes for a single user might evolve. For example, suppose there exists $N$ different dimensions of a user's preferences for i.e. food. The state $x_i(t)$ represents strength of preference for one of two extremes, and the preferences shape each other (i.e. if a user starts to prefer spicy food they also begin to prefer certain drinks). If the recommender system attempts select from a combinatorial set of meals to recommend, we can see how our dynamics can explore the feedback loops between user's tastes and the recommendations of the algorithm.

Either way, we can summarise the dynamics of the system as:

\begin{align}\label{eq:curtain}
    x_i(t+1) = (1-\alpha) \sum_j a_{ij} x_j(t) + \alpha \text{sign}(x_i(t)) \\
    \Rightarrow x(t+1) = (1-\alpha) A x(t) + \alpha \sigma(x(t))
\end{align}

Where $x(t) \in [-1,1]^N$ represents the tastes of each user, $\sigma(t) \in \{-1,1\}^N$ represents the possible configurations of each personalization algorithm, $A$ is a stochastic matrix representing the weights nodes place on neighbours, and $\alpha$ denotes the strength of the recommender influence. The dynamics conferred by this model are rich, and we do not go into a great amount of detail (see \cite{sikder2020minimalistic} for a more in-depth analysis), but the key aspect we are interested in are how models with private signals can support both heterogeneous and consensus outcomes depending on the parameterisation.

For example, if we take $\alpha > 0.5$, we can see that the sign of agents will never change, so the private signals will be fixed, and with high probability the resulting steady state $x^* = \alpha(I-(1-\alpha)A)^{-1}\sigma(x(0))$ will be heterogeneous. However, as $\alpha$ falls below $0.5$, the configurations of the recommendations will begin to vary over time. It turns out that as $\alpha$ falls (as the recommender effects get \textit{weaker}), it increases the probability of a cascade occurring, in which case the personalisation systems all begin to align in their recommendations ($\sigma(t) \to \pm \mathbbm{1}$) and as a result all agents end up with a consensus around $+1$ or $-1$. In other words, the weaker the recommender effects, the less diversity is promoted for users (in the form of lower expected heterogeneity). This is illustrated in Figure \ref{fig:curtain}.

\subsection{Linear quadratic games}

Finally we consider an example where update matrices are not dynamic ($A(t) = A$ and $b(t) = b$) but our random walk interpretations can provide useful intuition as to the distributions of the heterogeneous steady state outcomes that occur. One important class of stationary models in economics that are nested in our dynamics are linear quadratic games. They are commonly used to model strategic complementarity in games played over networks, and have been used to investigate empirical questions ranging from criminal activity to educational attainment to industrial organisation (see \cite{jackson2015games} for a review).

The basic setup is as follows\footnote{More general forms are possible, but incur a great deal of extra notation. For example, \cite{lambert2018quadratic} consider Linear Quadratic Gaussian games with multi-dimensional action spaces for agents alongside a learning framework where private rewards are realised stochastically, resulting in analysis for a Bayes-Nash equilibrium instead. We consider a one-dimensional, deterministic version for simplicity.}. For a set of $N$ agents, each agent $i$ chooses an effort level $x_i \geq 0$ that incurs a private reward $r_i x_i$ and a private cost $\frac{x_i^2}{2}$. Furthermore, the agent also receives a spillover $a_{ij}x_i x_j > 0$ reward from co-ordinating activity levels with other nodes $x_j$. Gathering this into a utility function we get:

\begin{equation}\label{eq:utility}
    U_i(x) = r_i x_i + \sum_{j} a_{ij} x_i x_j - \frac{x_i^2}{2}
\end{equation}

A typical example used is criminal activity, for example in \cite{ballester2010delinquent}. In such models, criminals choose a level of criminal activity to engage in. Criminal activity results in some expected private reward, which increases as more associates are involved in the crime. The costs can capture for example the probability of capture. We can see therefore that the utility structure encourages agents to engage in more activity the more of their peers do so.

Solve the partial derivative of \ref{eq:utility} with respect to $x_i$ provides us with the best reply dynamics:

\begin{align}
    x_i(x_{-i}) = \underset{x_i}{\text{argmax}} [U_i(x_i, x_{-i})] = \sum_j a_{ij} x_j + r_i \\
    \Rightarrow x(k+1) = Ax(k) + r
\end{align}

Where in the last step we just vectorised the best reply function to the states indexed at $k$ to return to our familiar affine form, with $A$ summarising interaction effects between agents and $r$ being a vector of private rewards. We can suppose the level of effort is bounded (i.e. infinite effort levels are ruled out, so the state space is compact). For the sake of exposition, we suppose the interaction effects matrix $A$ is row substochastic (i.e. $\sum_j a_{ij} < 1$)\footnote{This is not a particularly restrictive assumption and is in fact closely related to the conditions required for a Nash equilibrium to exist in this game, for example see \cite{ballester2006s}. It is straightforward to generalise from the example we consider.}. The best reply dynamics therefore converge to the Nash Equilibrium, which is, as expected:

\begin{align*}
    x^* = \underbrace{(I-A)^{-1}}_{F}\underbrace{r}_{B}
\end{align*}

We can see that the equilibrium outcomes of such games will in general be heterogeneous (this is almost sure if elements of $A$ are drawn from some independent continuous distribution and $r \neq c \mathbbm{1}$). This heterogeneity will be the case even if the private signal vector $r$ is not particularly varied - for example it can consist of only two levels of reward $r_1$ and $r_2$. The rows of the fundamental matrix $F = (I-A)^{-1}$ encode how small differences in topological position of the node in the weighted graph implicitly represented by $A$ will encourage different steady state actions are adopted by the agents. Put differently, even if the variation in private signals is low, the topological variation is often sufficient to induce heterogeneous outcomes where each agent adopts a unique strategy in equilibrium.

We can also use the analogy of random walks we have developed to build some useful intuition about the general characteristics of such equilibrium. For example, note that the partial derivative of an agent's steady state outcome with regards to their own private reward is:

\begin{equation}
    \frac{\partial x_i^*}{\partial r_i} = F_{ii}
\end{equation}

\begin{figure}
    \centering
    \includegraphics[width=0.6\textwidth]{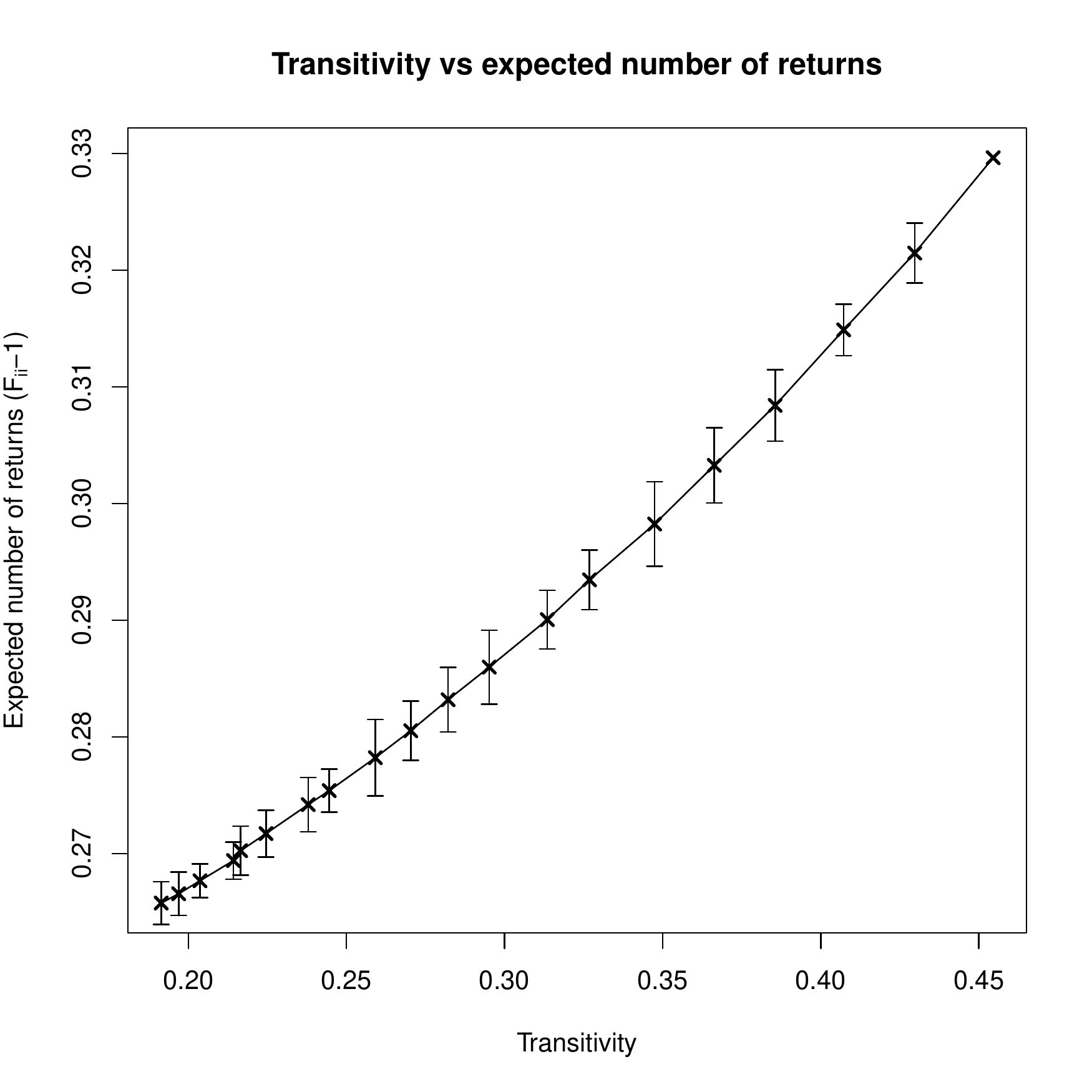}
    \caption{We generated a sequence of small-world networks rewired from a 2D lattice with nodes connected to neighbours up to distance 2 and $N = 100$. The rewiring probability ranges in 20 steps from 0.2 to 0 with 20 iterations in each case. The interaction matrix is realised with $a_{ij} = \frac{0.95}{k_i}$ if $i$ and $j$ are connected, where $k_i$ is the degree of $i$. The mean transitivity at each level of rewiring is displayed alongside the mean value of $F_{ii} - 1$ for all nodes $i$. We can see that as transitivity increases, the expected number of returns increase. Note the solid line merely connects the numerical means, and is used to emphasize the monotonic increase.}
    \label{fig:trans}
\end{figure}

Recall that the diagonal elements of the fundamental matrix encode the expected number of times a random walk that commences at $i$ hits $i$ before it is absorbed (the expected number of returns is $F_{ii} - 1$). Therefore, we can conclude that any change in the network topology that increases the number of cycles (while holding all other features fixed) will increase the attention agents pay to their own private rewards in equilibrium. We can sense check this in Figure \ref{fig:trans}, where we consider a range of small-world networks generated on a 2D lattice with a decreasing rewiring probability. As the rewiring probability $p$ decays to $0$, the network shifts from an Erdos-Renyi network with low transitivity to a lattice with high transitivity. The (mean) transitivity is measured for $p$ ranging from $0.2$ to $0$, and is compared to the mean of $F_{ii} - 1$, the expected number of returns of a random walk to each node, and the partial derivative we are interested in.

Therefore, we can see that as the underlying interactions become increasingly transitive, the strategic choices of each agent will be more heavily influenced by their private reward. Suppose for example we wished to modulate the expected level of activity of some agent by reducing their private reward (in the example of criminal networks, this can be translated as increasing surveillance on that agent, increasing their probability of capture and reducing their expected reward for activity). We can see this strategy will be more effective if the network is highly transitive. This occurs naturally if, for example, agent interactions are shaped by physical proximity, which might be the case for i.e. physical crimes as opposed to cyber-crimes. Intuitively, this occurs because increases in an agent's activity have a greater spillover effect if local clusters are closely connected and reinforce each other, whereas agents with disconnected neighbours will have less reinforcement between those neighbours.

\section{Conclusion}

In this paper, we have considered the problem of modelling heterogeneous outcomes in multi-agent systems. As demonstrated, many models on such systems will surely result in consensus, which is often unrepresentative of the real empirical phenomenon we are wishing to investigate. In order to address this, we developed a set of necessary criteria for our models to instead produce heterogeneous outcomes, where each agent possesses a unique outcome in the steady state. Furthermore, through an appropriate analogy with random walks on graphs, we provide an intuitive characterisation of the features of this steady state, which can help us ensure our desired model contains the features we may be hoping to represent in the real-world phenomenon.

One of the key insights from our analysis was that for strongly connected graphs $\mathcal{G}_\infty$, private signals were a necessary feature to ensure heterogeneous outcomes were possible. An intuitive way of seeing why this is the case follows from observing that the averaging dynamics enforced by the graph structure ($A(t)$) is an inherently convex operation, and by necessity ensures that the span of the agents as a result of averaging is contained in the convex hull of the original set of states. In isolation, the hull must shrink, leading to the consensus outcomes we are familiar with. The presence of private signals helps us break out of the convexity of these dynamics, and provide in some sense external perturbations that allow agents to explore the state space instead of iteratively compounding any similarity that exists between agents.

One shortcoming of our analysis is that we were not able to provide \textit{sufficient} criteria for our models to result in heterogeneous outcomes. Given the structure we know our outcomes must take ($F B$), any such theorem is likely to be related to the eigen-structure of the interaction matrices, and a precise measure-theoretic analysis for distributions over such matrices. We consider this a promising direction for future study.

\section*{Appendix}

\subsection*{A1: Private signals and ghost nodes}

We demonstrate how a model with private signals $b(t) \in \mathcal{X} \subset \mathbb{R}^{N \times d}$ as in Equation \ref{eq:affine} can be written in the augmented form:

\begin{align}\label{eq:auglin}
    \tilde{X}(t+1) = \tilde{A}(t)\tilde{X}(t)
\end{align}
Where $\tilde{A}(t)$ is a row stochastic matrix. We have made the following augmentations:

\begin{align}\label{eq:auglinear}
    \tilde{X}(t)  = 
        \left[
        \begin{array}{c}
        X(t) \\
        \hline
        C
        \end{array}
        \right] \in \mathbb{R}^{(N+2d)\times d} \\
    \tilde{A}(t) = 
        \left[
        \begin{array}{c | c}
        (I-\Lambda) A(t) & \Lambda W(t)\\
        \hline
        0 & I
        \end{array}
        \right] \in \mathbb{R}^{(N+2d)\times (N+2d)}
\end{align}

\begin{align}
    C  =
        d \left[
        \begin{array}{c|c|c|c}
        \overline{x}_1 & 0 & \ldots & 0\\
        \underline{x}_1 & 0 & \ldots & 0\\
        \hline
        0 & \overline{x}_2 & \ldots & 0\\
        0 & \underline{x}_2 & \ldots & 0\\
        \hline
        \ldots & \ldots & \ldots & \ldots \\
        \hline
        0 & 0 & \ldots & \overline{x}_d\\
        0 & 0 & \ldots & \underline{x}_d\\
        \end{array}
        \right] \in \mathbb{R}^{2d \times d}
\end{align}     

\begin{align}
    W(t) = 
        d^{-1} \left[
        \begin{array}{c c|c c|c|c c}
        \overline{w}^{(1)}_1(t) & \underline{w}^{(1)}_1(t) & \overline{w}^{(1)}_2(t) & \underline{w}^{(1)}_2(t) &
        \ldots &
        \overline{w}^{(1)}_d(t) & \underline{w}^{(1)}_d(t) \\
        \hline
        \overline{w}^{(2)}_1(t) & \underline{w}^{(2)}_1(t) & \overline{w}^{(2)}_2(t) & \underline{w}^{(2)}_2(t) &
        \ldots &
        \overline{w}^{(2)}_d(t) & \underline{w}^{(2)}_d(t) \\
        \hline
        \ldots & \ldots &
        \ldots & \ldots &
        \ldots &
        \ldots & \ldots \\
        \hline
        \overline{w}^{(N)}_1(t) & \underline{w}^{(N)}_1(t) & \overline{w}^{(N)}_2(t) & \underline{w}^{(N)}_2(t) &
        \ldots &
        \overline{w}^{(N)}_d(t) & \underline{w}^{(N)}_d(t) \\
        \end{array}
        \right] \in \mathbb{R}^{N \times 2d}
\end{align}
Where $\overline{x}_l$ is the upper bound of the $l$-th dimension of the state vector space $x$, and $\underline{x}_l$ for the corresponding lower bound\footnote{Recall that $X$ is a compact set, and is thus bounded}. $\overline{w}^{(i)}_l(t) +  \overline{w}^{(i)}_l(t) = 1$ then refer to the weights the $i$-th node places on the upper and lower bounds of the $l$-th component respectively, ensuring that we can express $b_{il}(t) = \overline{w}^{(i)}_l(t)\overline{x}_l + (1-\overline{w}^{(i)}_l(t))\underline{x}_l$, and more generally that $b(t) = W(t)C$. The preceding $d$ and $d^{-1}$ weights in front of the matrices merely ensure that the matrix $\tilde{A}(t)$ remains row stochastic. 

\subsection*{A2: Proof of Theorem 1 and Corollary 1}

Theorem \ref{thm:scc} states that for any model expressible as:

\begin{equation}
    x(t+1) = A(t) x(t)
\end{equation}

With $\mathcal{G}_\infty(\{A(t)\})$ defined as in the main text, all sink strongly connected components of $\mathcal{G}_\infty$ must converge to consensus, so long as the regularity assumptions are fulfilled. As a reminder, these are, for some $\delta > 0$:

\begin{enumerate}
    \item $A_{ij}(t) \in \{0\} \cup [\delta,1]$
    \item $A_{ii}(t) \geq \delta, \forall i$
\end{enumerate}

In order to prove our result we can utilize the following result from \cite{wolfowitz1963products}, which makes use of some extra terminology. Define $\gamma(A)$:

\begin{equation}
    \gamma(A) = \underset{j}\max\underset{i_1,i_2}{\max}|a_{i_1,j} - a_{i_2,j}|
\end{equation}

That is, $\gamma(A)$ measures the extent to which the rows of $A$ vary. Furthermore, a stochastic matrix $A$ is indecomposable and aperiodic (SIA) iff $A^* = \underset{t \to \infty}{\lim}A^t$ exists and $\gamma(A^*)=0$. Then the first theorem in \cite{wolfowitz1963products} states:

\begin{theorem}\label{thm:wolf}

For any product of stochastic matrix $A(t)A(t-1) \ldots A(1)A(0)$, let any subproduct (product of some subset of consecutive matrices) be SIA. Then for any $\epsilon>0$ there exists $n(\epsilon)$ such that any subproduct $A$ of length $n$ satisfies $\gamma(A)<\epsilon$.

\end{theorem}

We can now proceed. Firstly, designate some $t_0$ such that for all $t \geq t_0$, $A(t)[i,j]>0 \to (i,j) \in \mathcal{E}_{\infty}$. That is, after some long enough time period, all finite interactions will cease, and any edges that are instantiated in the $A(t)$ matrices must be drawn from the infinite edge set. Without loss of generality\footnote{All quasi-connected components can be partitioned into a block that does not interact with or otherwise influence the strongly connected components.}, assume that $\mathcal{G}_\infty$ consists only of $k$ SSCCs denoted $\mathcal{C} = \{C_1, C_2, \ldots, C_k\}$.

Importantly, this means there exists some permutation of $A(t)$ which organises the matrix into block diagonals, where each block diagonal is a stochastic matrix corresponding to a sink strongly connected component of $\mathcal{G}_{\infty}$ (that is, no paths exist in either direction between two components of $\mathcal{G}_{\infty}$; recall that in our definition of SSCCs there are no outgoing paths from each SSCC). In this case we can designate $x(t_0) = \prod_{t=0}^{t_0}A(t)x(0)$, and restart the dynamics with $x(t_0)$ as our new initial vector. We can now see that the matrix updates will be of the form:

\begin{equation}
    x(t+1) = 
    \left[
    \begin{array}{c c c}
    A_1(t) & \cdots & 0 \\
    \vdots & \ddots & \vdots \\
    0 & \cdots & A_k(t)
    \end{array}
    \right]
    x(t)
    =
    \left[
    \begin{array}{c c c}
    \prod_{n=t_0}^{t}A_1(n) & \cdots & 0 \\
    \vdots & \ddots & \vdots \\
    0 & \cdots & \prod_{n=t_0}^{t}A_k(n)
    \end{array}
    \right]
    x(t_0)
    \end{equation}

Establishing the asymptotic properties of this process simplifies to establishing the asymptotic properties of $\prod_{n=t_0}^{t}A_r(n)$, since the blocks do not otherwise interact.

The simplest case arises when the block is invariant ($A_r(n) = A_r, \forall n \geq t_0$). In this case, all known results about DeGroot models can be applied directly (see, for example, \cite{proskurnikov2017tutorial}). In particular, by Assumptions (2), the subgraph $\mathcal{G}_\infty^{(r)}$ is strongly connected and aperiodic. Then $A_r$ is irreducible and $A_r^n \to \mathbbm{1}w'$, meaning that the nodes in this block will converge to consensus.

A more general case is when $A_r(n)[i,j] > 0 \iff (i,j) \in \mathcal{E}_\infty$. That is, each (sub)matrix contains \textit{all} (as opposed to a subset of) the edges of the infinite interaction matrix, but the weights may vary. Denoting as $\hat{A}_r$ the ``mean'' interaction (sub)matrix, the stochastic matrix induced by row normalizing the adjacency matrix of the infinite (sub)graph ($\hat{A}_r = A[\mathcal{G}_\infty^{(r)}]$). The product $\prod_{n=t_0}^{t}A_r(n)$ will inherit many of the properties of $\hat{A}_r^{(t-t_0+1)}$. In particular we can prove the following Lemma:

\begin{lemma}
   For some graph $\mathcal{G}$, let $A[\mathcal{G}]$ be an induced adjacency matrix where $A_{ij} \geq \delta \iff (i,j) \in \mathcal{E}(\mathcal{G})$ for some $\delta > 0$. Similarly, let $\mathcal{G}[A]$ be the graph induced by a square matrix $A$. Consider some arbitrary finite set of $p$ induced adjacency matrices $\{A(1)[\mathcal{G}], A(2)[\mathcal{G}], \ldots, A(p)[\mathcal{G}]\}$ and arbitrary reference matrix $\hat{A} = A[\mathcal{G}]$. Then $\tilde{\mathcal{G}}(\hat{A}^p) = \tilde{\mathcal{G}}(\prod_{n=0}^p A(n))$.
\end{lemma}

Put differently, if we take some arbitrary reference matrix generated from a graph ($\hat{A}[\mathcal{G}]$) and raise it to some power $p$, we can denote the graph induced by $\hat{A}^p$ as $\tilde{\mathcal{G}}(\hat{A}^p)$. If we do the same with a a product of adjacency matrices $\prod_{n=0}^p A(n)$ where each $A(n)$ also fully realises the original graph $\mathcal{G}$, the resulting product will generate the same graph $\tilde{\mathcal{G}}$.

This follows from induction. For the base case, note that under the definition we can directly see $\mathcal{G}(\hat{A}) = \mathcal{G}(A(n))$.  Now suppose it holds for some arbitrary $(m-1)$. Then for $m$, we can see:
\begin{align}
    (i,j) \in \mathcal{E}(\mathcal{G}(\hat{A}^m)) \iff \hat{A}^m[i,j] > 0 \\
    \iff \langle \hat{A}[i,], \hat{A}^{m-1}[,j] \rangle > 0 \\
    \iff \exists (i,k) \in \mathcal{E}(\mathcal{G}(\hat{A})) \land (k,j) \in \mathcal{E}(\mathcal{G}(\hat{A}^{(m-1)}))
\end{align}
But note that since by the inductive step, $\mathcal{G}(\hat{A}^{(m-1)}) = \mathcal{G}(\prod_{n=0}^{m-1}A(n))$. Together with the base case:
\begin{align}
    \exists (i,k) \in \mathcal{E}(\mathcal{G}(\hat{A})) \land (k,j) \in \mathcal{E}(\mathcal{G}(\hat{A}^{(m-1)})) \\
    \iff \exists (i,k) \in \mathcal{E}(\mathcal{G}(A(m))) \land (k,j) \in \mathcal{E}(\mathcal{G}(\prod_{0}^{m-1}A(n))) \\
    \iff \langle A(m)[i,], [\prod_{0}^{m-1}A(n)][,j] \rangle > 0 \\
    \iff \prod_{0}^{m}A(n)[i,j] > 0 \\
    \iff (i,j) \in \mathcal{E}(\mathcal{G}(\prod_{0}^{m}A(n))\\
    \Rightarrow (i,j) \in \mathcal{E}(\mathcal{G}(\hat{A}^m)) \iff (i,j) \in \mathcal{E}(\mathcal{G}(\prod_{0}^{m}A(n)) \\ 
    \Rightarrow \mathcal{G}(\hat{A}^m) =  \mathcal{G}(\prod_{0}^{m}A(n))
\end{align}

Giving us our desired result. We can therefore proceed with the knowledge that the graph induced by any power of the (sub)matrix $\hat{A}_r^p$ will be identical to the graph induced by the product of the $p$ terms $\prod_p A_r(n)$. That is, $\mathcal{G}[\hat{A}_r^p] = \mathcal{G}[\prod_p A_r(n)]$. Importantly, this means that any properties that are inherited by any adjacency matrix of such a graph are equivalent between these two representations.

To exploit this property, note that if the (sub)graph $\mathcal{G}_\infty^{(r)}$ is strongly connected and aperiodic, then so is $\tilde{\mathcal{G}}[\hat{A}_r[\mathcal{G}_\infty^{(r)}]^p]$.\footnote{$\mathcal{G}_\infty^{(r)}$ is strongly connected and aperiodic if and only if its adjacency matrix $A[\mathcal{G}_\infty^{(r)}]$ is primitive (see, for example, Section 1.3 of \cite{levin2017markov} for a discussion). This means that for some power $q$, $A[\mathcal{G}_\infty^{(r)}]^q$ has only strictly positive entries. This means that any powers of the matrix are also primitive, and therefore the graph induced by this power matrix must also possess strong connectivity and aperiodicity.} Since $\mathcal{G}(\hat{A}_r^p) = \mathcal{G}(\prod_{n=t_0}^{t}A_r(n))$, the graph induced by the product of any set of consecutive matrices $A_r(n)$ is also strongly connected and aperiodic. Finally, this means that the product of the matrices themselves,  $\prod_{n=t_0}^{t}A_r(n)$ are stochastic, irreducible and aperiodic. Since this holds for any $p$ we therefore fulfil the conditions of Theorem \ref{thm:wolf}, and as such we can see the limit of the products of these matrices is a consensus matrix. That is, $\underset{t \to \infty}{\lim}\prod_{n=t_0}^{t}A_r(n) \to \mathbbm{1}a_r'$.

Finally, consider the most general case where there are no restrictions on $A_r(n)$ (except of course that edges are only drawn from the infinite graph). By Assumption 2, $A_r(n)[i,i]\geq\delta$ for all $i$, then it is straightforward to see that the product of any two consecutive matrices $AB$ will contain the edges of both the matrices. Since all edges in the infinite graph must recur, we can always partition the product $\underset{t \to \infty}{\lim}\prod_{n=t_0}^{t}A_r(n)$ into subproducts where each subproduct ``hits'' all the edges from the infinite graph, ensuring that the subproduct contains all the edges from the infinite graph. Now we are simply in the regime where each matrix is a (full) realisation of the infinite graph, and the results from above apply.

Corollary \ref{cor:scc} follows straightforwardly from the definition of heterogeneity.

\subsection*{A3: Proof of Theorem 2, Corollary 2, Corollary 3 and Corollary 4}

As a reminder, Theorem 2 states:
\setcounter{theorem}{1}
\begin{theorem}
Suppose Assumptions (1) and (2) hold for $t \geq t_0$. Let edges between quasi-connected nodes, SSCCs and quasi-connected nodes to SSCCs on $\mathcal{G}_\infty$ be represented in each $A(t)$ by $Q(t)$, $S(t)$ and $R(t)$ respectively. Then the model converges if and only if for all quasi-connected nodes one of the following conditions is met:
\begin{itemize}
    \item $Q(t) \to Q$ and $R(t) \to R$
    \item $R(t)S = (I-Q(t))M + \epsilon(t)$
\end{itemize}
Where $\epsilon(t) \to 0$, $M$ is an arbitrary row-stochastic matrix, and $S = \underset{t\to \infty}{\lim}S(t:t_0)$.
\end{theorem}

The first thing to note is that the first condition of Theorem \ref{thm:converge}, ($Q(t) \to Q$, $R(t) \to R$) is a special case of of the second condition, since we can always set $M = (I-Q)^{-1}RS$ (Recall that since $S(t)$ consists of sink strongly connected components with positive self-weights, $S = \underset{t\to \infty}{\lim}S(t:t_0)$ is well-defined as per Theorem \ref{thm:scc}). Then, the left hand expression converges to $RS$. The right hand expression converges to:

\begin{equation}
  \underset{t \to \infty}{\lim}(I-Q(t))^{-1}(I-Q)RS = (I-Q)^{-1}(I-Q)RS = RS  
\end{equation}

Therefore, the burden of proof is on the second condition:
\begin{itemize}
    \item \textit{$R(t) S = (I-Q(t)) M + \epsilon(t)$, where $S = \lim \prod S(t)$, $M$ is some stochastic matrix, and $\epsilon(t) \to 0$.}
\end{itemize}

Which clarifies the sole, highly specific condition where the dynamics can converge without the convergence of each sub-matrix.
Recall that for $t \geq t_0$, $A(t) = \bigl( \begin{smallmatrix}Q(t) & R(t)\\ 0 & S(t)\end{smallmatrix}\bigr)$, where $Q(t) \in \mathbb{R}^{m\times m}$, $R(t) \in \mathbb{R}^{m\times p}$, $S(t) \in \mathbb{R}^{p\times p}$ and $n = m+p$. Let us denote matrix products as $M(t_1:t_0) = \prod_{n=t_0}^{t_1} M(n)$. The state vector $x(t)$ can then be written:
\begin{align}
    x(t+1) = 
    \left[
    \begin{array}{c | c}
    Q(t) & R(t) \\
    \hline
    0 & S(t)
    \end{array}
    \right]
    x(t)
    =
    \left[
    \begin{array}{c | c}
    Q(t:t_0) & R(t:t_0) \\
    \hline
    0 & S(t:t_0)
    \end{array}
    \right]
    x(t_0)
\end{align}
From here we can draw some quick conclusions. Note that $S(t:t_0)$ just consists of block diagonal sink strongly connected components (each absorbing set), and therefore by Theorem \ref{thm:scc}, $S(t:t_0) \to S$, where $S$ consists of block diagonal consensus matrices.

Next, $Q(t:t_0) = \prod_{n=t_0}^{t}Q(n) \to 0$. Let $\mathcal{G}_\infty^{[Q]}$ denote the subgraph of the infinite graph consisting of these nodes. Let the nodes of this subgraph that are directly connected to an SSCC be called the ``exit nodes''. Denote the $i$-th rowsum of a matrix $M(t)$ as $\norm{M_i(t)}$.

We will prove the follow specific claim. Start from any $t^* \geq t_0$. For any node $i \in \mathcal{G}_\infty^{[Q]}$ at distance $d$ from an exit node, there exists some $t^{(d)} \geq t^*$ such that $\norm{Q_i(t:t^*)} \leq (1-\delta^{T (d+1)})<1$ for all $t \geq t^{(d)}$. Here, $T \geq 0$ is the longest time between which all edges of the infinite graph $\mathcal{G}_\infty$ are realised at least once.



The proof follows by induction. Let us begin with $d=0$ (i.e. for any exit node). We pick some arbitrary $t^* \geq t_0$ to begin our analysis. There will exist some $t^{(0)} \geq t^*$ such that an edge from the exit node $i$ to the SSCC $c$ is activated, in which case it must be realised with weight $a_{ic} \geq \delta$. As such, the $i$-th rowsum $\norm{Q_i(t^{(0)})} \leq (1-\delta)$ (since the matrix $A(t^{(0)})$ is row stochastic).  For the next time step $t^{(0)}+1$, we can see that:

\begin{align}
   \norm{Q_i(t^{(0)}+1:t^{(0)})} = Q_{ii}(t^{(0)}+1)  \norm{Q_i(t^{(0)})} + \sum_{j \neq i}Q_{ij}(t^{(0)} + 1)\underbrace{\norm{Q_j(t^{(0)})}}_{\leq 1}   \\
   \leq \underbrace{Q_{ii}(t^{(0)}+1)}_{\geq \delta}  (1-\delta) + \sum_{j \neq i}Q_{ij}(t^{(0)}+1) \leq \delta(1-\delta) + (1-\delta) < 1
\end{align}

We can repeat the above argument until $t^{(0)} + T$ to show that:

\begin{align}
   \norm{Q_i(t^{(0)}+T:t^{(0)})} \leq (1-\delta)(1 + \delta + \delta^2 + \ldots + \delta^{T}) = 1-\delta^{T} < 1
\end{align}

Since the edge $a_{ic}$ must be realised by $T$, the rowsum bound would be ``reset'' back to $(1-\delta) < (1-\delta^{T})$. This latter quantity therefore represents an upper bound for all exit nodes for $t \geq t^{(0)}$.

Now suppose the claim holds for any $d$. Consider a node $i$ at distance $d+1$. We know that for all $t \geq t^{(d)}$, the rowsum $\norm{Q_k(t:t^{(0)})}$ of their neighbour $k$ will be upper bounded by $(1-\delta^{(d+1)T})$. Suppose the next time the edge to the neighbour is realised is $t^{(d)} + T \geq t^{(d+1)} > t^{(d)}$. Since the edge $Q_{ik}(t^{(d+1)})$ is realised with value at least $\delta$, we can see that:

\begin{align}
   \norm{Q_i(t^{(d)}:t^{(0)})} = Q_{ik}(t^{(d+1)})  \underbrace{\norm{Q_k(t^{(d+1)}-1:t^{(0)})}}_{\leq (1-\delta^{(d+1)T})} + \sum_{j \neq k}Q_{ij}(t^{(d+1)})\underbrace{\norm{Q_j(t^{(d+1)}-1:t^{(0)})}}_{\leq 1}   \\
   \leq \underbrace{Q_{ik}(t^{(d+1)}) }_{\geq \delta}  (1-\delta^{(d+1)T}) + \sum_{j \neq k}Q_{ij}(t^{(d+1)}) \leq \delta(1-(1-\delta^{(d+1)T})) + (1-\delta) < 1
\end{align}

We now repeat the steps for the base case to get:

\begin{align}
   \norm{Q_i(t^{(d+1)}+T:t^{(0)})} \leq (1-\delta)(1 + \delta + \delta^2 + \ldots + \delta^{T-1}) + \delta^{T}(1-\delta^{(d+1)T}) \\
   = (1-\delta^T) + \delta^T - \delta^{(d+2)T} = 1-\delta^{(d+2)T}< 1
\end{align}

Once again, since all edges are realised by $T$ steps, we can conclude for the node $i$ at distance $(d+1)$ from an exit node, for all $t \geq t^{(d+1)}$, the row sum $\norm{Q_i(t:t^{(0)})} \leq (1-\delta^{(d+2)T})$. Suppose the longest path from an exit node to all nodes in $\mathcal{G}_\infty^{[Q]}$ is $D$. We can conclude therefore that for all $t \geq t^{(D)}$, the row sum $\norm{Q_i(t:t^*)} \leq \norm{Q_i(t:t^{(0)})} \leq (1-\delta^{(D+1)T}) = (1-\gamma)$ for all quasi-connected $i$.

Recall that $t^{(d)} \leq t^{(d-1)}+T$. Therefore, $t^{(D)} \leq t^* + (D+1)T$. It follows therefore that for any starting point $t^*$, we can conclude that $\norm{Q((t^* + (D+1)T):t^*)}_infty \leq (1-\gamma)$. Here $\norm{M}_\infty$ is the maximum row sum for the matrix $M$.

Finally, we note that $\underset{t\to\infty}{\lim} Q(t:t_0) = \prod_{n=t_0}^\infty Q(n)$ can be partitioned into subproducts of length $(D+1)T$, which we refer to as $\tilde{Q}(n)$. Each subproduct will have $\norm{\tilde{Q}(n)} \leq (1-\gamma)$. In the following let $\norm{M} = \norm{M}_\infty$:

\begin{align}
    \underset{t\to\infty}{\lim} \norm{Q(t:t_0)} = \underset{t\to\infty}{\lim} \norm{\prod_{n=0}^t \tilde{Q}(n)} \leq \underset{t\to\infty}{\lim} \prod_{n=0}^t \norm{\tilde{Q}(n)} \leq \underset{t\to\infty}{\lim} (1-\gamma)^t = 0 \\
    \Rightarrow     \underset{t\to\infty}{\lim} Q(t:t_0) = 0
\end{align}

Since the other blocks are guaranteed to converge, in order to show that $A(t:t_0)$ converges, we just need to prove that $R(t:t_0)$ converges under the conditions we stated. Beginning in the easier direction (only if), firstly suppose $R(t:t_0) \to M$. Then the steady state is just:

\begin{equation}
    x^* 
    = \underset{t \to \infty}{\lim} x(t)
    = \underset{t \to \infty}{\lim} A(t:t0) x(t_0)
    =  \left[
    \begin{array}{c | c}
    0 & M \\
    \hline
    0 & S
    \end{array}
    \right]
    x(t_0)
\end{equation}

Clearly, since $M$ is a block in the stochastic matrix $\prod_{n=t_0}^\infty A(n)$, it must also be stochastic. We want to show that $R(t) S \to (I-Q(t))M$. The matrix $R(t:t_0)$ updates as follows:
\begin{align}
    R(t:t_0) = Q(t) R(t-1:t_0) + R(t) S(t-1:t_0)\\
    \Rightarrow \underset{t\to\infty}{\lim} R(t:t_0) = \underset{t\to\infty}{\lim} (Q(t) R(t-1:t_0) + R(t) S(t-1:t_0)) \\
    = \underset{t\to\infty}{\lim} (Q(t) M + R(t) S + Q(t) \epsilon(t)^{(M)} + R(t) \epsilon(t)^{(S)})) \\
    = \underset{t\to\infty}{\lim} (Q(t) M + R(t) S) + \underset{t\to\infty}{\lim}(Q(t) \epsilon(t)^{(M)} + R(t) \epsilon(t)^{(S)}) \\
    = \underset{t\to\infty}{\lim} (Q(t) M + R(t) S)\\
    \therefore \underset{t\to\infty}{\lim} (Q(t) M + R(t) S) =  \underset{t\to\infty}{\lim} R(t:t_0) = M
\end{align}
In the third step we introduced $\epsilon(t)^{(M)} = M - R(t-1:t_0)$ (and analogously for $S$) where we know $\norm{\epsilon(t)^{(M)}} \to 0$. This allows us to split the limit without issue in the fourth step. In order to complete this direction of the proof we just need to define appropriate terms. Define $\epsilon(t) = (Q(t) M + R(t) S - M)$. We can see that $\underset{t\to\infty}{\lim} (\epsilon(t)) = \underset{t\to\infty}{\lim}(Q(t) M + R(t) S) - M = M-M= 0$. Then we can just re-arrange to obtain:
\begin{align}
    R(t) S= (I - Q(t)) M + \epsilon(t)
\end{align}
Where $\underset{t\to\infty}{\lim} (\epsilon(t)) = 0$, the result we wanted.

Now we prove the other direction (if). Suppose $R(t) S = (I-Q(t)) M + \epsilon(t)$  for some $M \in \mathbb{R}^{(m \times p)}$ where $M_{ij}\geq 0, M\mathbbm{1} = \mathbbm{1}$ (i.e. $M$ is row stochastic). Then:
\begin{align}
    R(t:t_0) = Q(t) R(t-1:t_0) + R(t) S(t-1:t_0)\\
    = Q(t) R(t-1:t_0) + R(t) S + R(t) \epsilon(t)^{(S)}\\
    = Q(t) R(t-1:t_0) + (I-Q(t))M + \epsilon(t) + R(t) \epsilon(t)^{(S)} \\
    = Q(t) (R(t-1:t_0) - M) + M + \underbrace{\epsilon(t) + R(t) \epsilon(t)^{(S)}}_{\delta(t)} \\
    =  Q(t) (Q(t-1) (R(t-2:t_0) - M) + \cancel{M} + \delta(t-1) - \cancel{M}) + M + \delta(t)
\end{align}
\begin{align}
= M + \underbrace{\prod_{t} Q(t) (R_0 - M)}_{\to 0} + \underbrace{(\delta(t) + Q(t) \delta(t-1) + Q(t) Q(t-1) \delta(t-2) + \ldots)}_{=\Delta(t)}
\end{align}
Finally we need to show that $\Delta(t) \to 0$, since we know $\delta(t) \to 0$. This is not entirely straightforward since the error terms $\delta(t)$ can in principle accumulate instead of diminishing exponentially (there is no guarantee $\norm{Q(t)} < 1$ for all $t$). In order to get around this note:

\begin{align*}
    \Delta(t) = Q(t) \Delta(t-1) + \delta(t) \\
    \Rightarrow \Delta(t + (D+1)T) = \prod_{n=t}^{(D+1)T+t} Q(n) \Delta(t-1) + (\prod_{n=t+1}^{(D+1)T+t} Q(n)\delta(t) + \prod_{n=t+2}^{(D+1)T+t} Q(n)\delta(t+1) + \ldots \\
    \ldots + \delta(t+(D+1)T)) \\
    \Rightarrow \norm{\Delta(t + (D+1)T)} \leq \underbrace{\norm{\prod_{n=t}^{(D+1)T+t} Q(n)}}_{\leq (1-\gamma)}\norm{\Delta(t-1)} + (\underbrace{\norm{\prod_{n=t+1}^{(D+1)T+t} Q(n)}}_{\leq 1}\norm{\delta(t)} + \ldots + \norm{\delta(t+(D+1)T)} \\
    = (1-\gamma) \norm{\Delta(t-1)} + \underbrace{(D+1)T \underset{n \in \{t, t+1, \ldots t+(D+1)T\}}{\max}\norm{\delta(n)}}_{\mu(t-1)} \\
    \Rightarrow \norm{\Delta(t + T_0)} \leq (1-\gamma)\norm{\Delta(t)} + \mu(t)
\end{align*}


Note that since $\norm{\delta(t)} \to 0$, the extraneous term $\mu(t)$ can be made arbitrarily small for large enough $t$. In order to ease the analysis consider the subsequence $(t, t+T_0, t+2T_0, \ldots) \to (k, k+1, k+2, \ldots)$. Therefore, we can rewrite the above as:

\begin{equation}
    \norm{\Delta(k+1)} \leq (1-\gamma)\norm{\Delta(k)} + \mu(k)
\end{equation}

We will show that each subsequence $\{\norm{\Delta(k)}\}$ converges to zero (and therefore the full sequence converges to zero). Now suppose by contradiction that $\liminf{\norm{\Delta(k)}} = \phi > 0$. We can pick some $k_0$ such that for all $k \geq k_0$, $\mu(k) \leq (1-\alpha) \phi$ for $(1-\gamma) < \alpha < 1$. It follows therefore that there always exists some $k$ such that $\norm{\Delta(k)} \leq \alpha \frac{\phi}{(1-\gamma)}$. Therefore, $\norm{\Delta(k+1)} = \alpha \phi + \epsilon(k) \leq \alpha \phi + (1-\alpha)\phi < \phi$. This means $\phi$ can no longer be the $\liminf$ of the sequence $\norm{\Delta(k)}$, leading to a contradiction. Therefore, $\liminf{\norm{\Delta(k)}} = 0$.

Finally, for completeness suppose $\limsup \norm{\Delta(k)} = \psi > 0$. Choose some $k_0$ and $\alpha < 1$ such that $\epsilon(k_0) \leq \alpha \gamma \psi$ for all $k \geq k_0$. Since $\liminf \norm{\Delta(k)} = 0$, there exists some $k \geq k_0$ such that $\norm{\Delta(k_0)} \leq \alpha \psi$. Then $\norm{\Delta(k_0+1)} \leq \alpha \psi (1-\gamma) + \alpha \psi\gamma  = \alpha \psi$. Since $\norm{\epsilon(k)} \leq \alpha \psi \gamma$ for all $k\geq k_0$, we can see that $\norm{\Delta(k)}$ is bounded above by $\alpha \psi$, and therefore $\limsup \neq \psi$, leading to a contradiction. Therefore, $\lim \norm{\Delta(k)} = \limsup \norm{\Delta(k)} = \liminf \norm{\Delta(k)} = 0$.

Since this occurs for any arbitrary subsequence, we can conclude that $\lim \norm{\Delta(t)} = 0$. Finally, we can conclude that $R(t:t_0) \to M$, and as such the entire process converges, proving Theorem \ref{thm:converge}.

Corollary \ref{cor:ss} follows from the block structure:



\begin{align}
    x(t+1) = 
   \left[
    \begin{array}{c}
    x_{QC}(t+1) \\
    \hline
    x_{SC}(t+1)
    \end{array}
    \right] =
    \left[
    \begin{array}{c | c}
    Q(t:t_0) & R(t:t_0) \\
    \hline
    0 & S(t:t_0)
    \end{array}
    \right]
    \left[
    \begin{array}{c}
    x_{QC}(t_0) \\
    \hline
    x_{SC}(t_0)
    \end{array}
    \right]
    \\ 
    \Rightarrow x(t) = 
    \left[
    \begin{array}{c}
    x_{QC}(t) \\
    \hline
    x_{SC}(t)
    \end{array}
    \right] \to
    \left[
    \begin{array}{c | c}
    0 & M \\
    \hline
    0 & S
    \end{array}
    \right]
    \left[
    \begin{array}{c}
    x_{QC}(t_0) \\
    \hline
    x_{SC}(t_0)
    \end{array}
    \right] \\
    \Rightarrow x_{QC}(t) \to M x_{SC}(t_0) \\
    \Rightarrow x_{SC}(t) \to S x_{SC}(t_0)
\end{align}

Corollary \ref{cor:convergeprivate} follows immediately from Theorem \ref{thm:converge} with the appropriate mapping of the augmented matrix to the general form illustrated above, noting in particular that $S(t:t_0) = I$ for all $t$.

One final point to make is that Corollary \ref{cor:convergeprivate} assumes that the infinite graph of the original nodes $\mathcal{G}_\infty$ is strongly connected for simplicity. Note that the results can be extended to a general case where $\mathcal{G}_\infty$ is quasi-connected, but we must make use of an additional Assumption: all nodes must be path connected to a node where $\lambda_i > 0$. To see this note that the proof for Theorem \ref{thm:converge} the convergence of $Q(t:t_0) \to 0$ makes use of the fact that all quasi-connected nodes were path connected to an ``exit node'' (if they were not, they would not be a quasi-connected node). The exit nodes in that proof correspond to nodes in a private signal model that place some non-zero weight on their private signals (i.e. $\lambda_i > 0$).

Corollary \ref{cor:hetero} states that heterogeneous steady states require that $A(t) \to A$ and $b(t) \to b$, or $A(t) \not{\to} A$ and $b(t) \not{\to} b$. In order to show this we rule out heterogeneous outcomes when one update matrix converges and the other does not.

Consider first if $A(t) \to A$. By Theorem \ref{thm:converge}, we know that for a convergent model (where $\lambda_i > 0 \forall i$):

\begin{align}
    \Lambda W(t) = (I - (I-\Lambda)A(t))M + \epsilon(t) \\
    \Rightarrow W(t) = \underbrace{\Lambda^{-1}(I-(I-\Lambda)A)M}_{L} + \delta(t)
\end{align}

Since $\delta(t) \to 0$, then $W(t) \to L$, meaning $W(t)C = b(t) \to b$. If $\lambda_i = 0$ for any $i$, then we can simply replace the $i$-th row of $W(t)$ with zeroes and let $\lambda_i = 1$ to repeat the above. As such, there cannot be an outcome where $A(t) \to A$ and $b(t) \not{\to} b$.

If $b(t) \to b$, then $W(t) \to W$, and we can see similarly by re-arranging the expression in Theorem \ref{thm:converge} that:

\begin{align}
    A(t) M = (I-\Lambda)^{-1}(M - \Lambda W) + \phi(t) \\
    \Rightarrow A(t) M C  = A(t) x^* =  \underbrace{(I-\Lambda)^{-1}(M - \Lambda W)C}_{K} + \kappa(t) = K(t)
\end{align}

Where in the second line we just made use of the generic steady state structure from Corollary \ref{cor:scc} to show that $x^* = MC$. Since $\kappa(t) \to 0$, it follows that $A(t)x^* = K(t) \to K$. Without loss of generality suppose that $d=1$ (i.e. $x_i$ is one-dimensional). We can write the $i$-th entry $K_i(t)$ as:

\begin{equation}
    K_i(t) = \sum_{j \in \mathcal{N}(i)} A_{ij}(t) x^*_j - \kappa_i(t)
\end{equation}

Suppose by contradiction that $\mathcal{H}(x^*) > 0$, in which case all pairs $x^*_i \neq x^*_j$. For any $i$, this means that the summand $\sum_{j \in \mathcal{N}(i)} A_{ij}(t)$ will vary whenever an edge $A_{ij}(t)$ varies. Since $\kappa_i(t) \to 0$, it follows that the term $K_i(t)$ will never converge, leading to a contradiction.




\bibliographystyle{unsrt}  
\bibliography{references}

\end{document}